\documentclass[amssymb,amsfonts,amsmath,reprint,aps,prapplied,floatfix]{revtex4-1}
\usepackage{graphicx}
\usepackage{hyperref}

\newcommand{\mrm}[1]{\mathrm{#1}}

\newcommand{\ketbra}[2]{| #1 \rangle\langle #2 |}
\newcommand{\expval}[1]{\langle #1 \rangle}
\newcommand{\abs}[1]{\left| #1 \right|}

\newcommand{\ext}{\mathrm{ext}}

\begin{document}

\title{A Nonlinear Charge- and Flux-Tunable Cavity Derived from an Embedded \\Cooper Pair Transistor}
\date{\today}

\author{B.~L.~Brock}
\email{Benjamin.L.Brock.GR@dartmouth.edu}
\affiliation{Department of Physics and Astronomy, Dartmouth College, Hanover, New Hampshire 03755, USA}

\author{Juliang Li}
\altaffiliation[Present address: ]{High Energy Physics Divison, Argonne National Laboratory, 9700 South Cass Avenue, Argonne, IL 60439, USA}
\affiliation{Department of Physics and Astronomy, Dartmouth College, Hanover, New Hampshire 03755, USA}

\author{S.~Kanhirathingal}
\affiliation{Department of Physics and Astronomy, Dartmouth College, Hanover, New Hampshire 03755, USA}

\author{B.~Thyagarajan}
\affiliation{Department of Physics and Astronomy, Dartmouth College, Hanover, New Hampshire 03755, USA}

\author{William~F.~Braasch~Jr.}
\affiliation{Department of Physics and Astronomy, Dartmouth College, Hanover, New Hampshire 03755, USA}

\author{\\ M.~P.~Blencowe}
\affiliation{Department of Physics and Astronomy, Dartmouth College, Hanover, New Hampshire 03755, USA}

\author{A.~J.~Rimberg}
\email{Alexander.J.Rimberg@dartmouth.edu}
\affiliation{Department of Physics and Astronomy, Dartmouth College, Hanover, New Hampshire 03755, USA}

\begin{abstract}
We introduce the cavity-embedded Cooper pair transistor (cCPT), a device which behaves as a highly nonlinear microwave cavity whose resonant frequency can be tuned both by charging a gate capacitor and by threading flux through a SQUID loop.  We characterize this device and find excellent agreement between theory and experiment.  A key difficulty in this characterization is the presence of frequency fluctuations comparable in scale to the cavity linewidth, which deform our measured resonance circles in accordance with recent theoretical predictions [Brock et al., Phys.~Rev.~Applied 14, 054026 (2020)].  By measuring the power spectral density of these frequency fluctuations at carefully chosen points in parameter space, we find that they are primarily a result of the $1/f$ charge and flux noise common in solid state devices.  Notably, we also observe key signatures of frequency fluctuations induced by quantum fluctuations in the cavity field via the Kerr nonlinearity.
\end{abstract}

\maketitle

\section{Introduction}

Tunable microwave cavities have found wide-ranging applications in recent years.  Charge-tunable cavities, for example, have been used as electrometers \cite{Sillanpaa2004, Bell2012} and as platforms for optomechanics \cite{Heikkila2014, Pirkkalainen2015}.  Flux-tunable cavities, on the other hand, have been used as parametric oscillators to achieve single-shot readout of superconducting qubits \cite{Lin2014, Krantz2016} and as platforms for studying the dynamical Casimir effect \cite{Wilson2011,Laehteenmaeki2013,Svensson2018}.  Since this tunability is often achieved via embedded Josephson junctions (JJs), which are inherently nonlinear, these cavities typically have significant nonlinearities that can be either helpful \cite{Yurke2006, Castellanos-Beltran2007, Tosi2019} or harmful \cite{Wustmann2013, Liu2017} depending on the application.  

Here we introduce the cavity-embedded Cooper pair transistor (cCPT), a device which consists of a Cooper pair transistor whose source and drain electrodes are connected between the voltage antinode of a quarter-wavelength microwave cavity and the ground plane, as shown in Fig. \ref{fig:sample_and_schematic}.  The phase coordinates of the Josephson junctions (JJs) that comprise the CPT couple to the cavity flux coordinate via their shared SQUID loop, such that the ground state energy of the CPT contributes to the effective potential of the cavity.  Since this ground state energy can be tuned both by charging the capacitance that gates the island of the CPT and by threading flux through the SQUID loop, so too can the resonant frequency be tuned by these parameters.  As in comparable devices, strong nonlinearities are induced in the effective cavity Hamiltonian due to the inherent nonlinearity of the JJs.  As a standalone device the cCPT can be used as an electrometer \cite{Brock2021_electrometry,Kanhirathingal2020}, magnetometer, parametric amplifier, and parametric oscillator.  One distinguishing feature of this device is the ability to combine electrometry with parametric pumping, which could be used for novel charge detection schemes analogous to those that have been used for the readout of superconducting qubits \cite{Lin2014,Krantz2016}.  Furthermore, the cCPT is the first building block of a promising scheme for achieving ultra-strong optomechanical coupling at the single-photon level \cite{Rimberg2014}.  

In the present work we focus on characterizing the cCPT as a function of our gate and flux parameters.  Measuring the tunable resonant frequency of the cCPT is straightforward, and we find excellent agreement between our measurements and our theoretical model; however, extracting the damping rates of the cCPT is complicated by the presence of frequency fluctuations comparable in scale to the cavity linewidth.  These fluctuations arise due to charge and flux noise coupling into the resonant frequency via its tunability, and quantum fluctuations of the cavity field coupling into the resonant frequency via the Kerr nonlinearity.  As recently predicted \cite{Brock2020}, such fluctuations deform the ideal resonance circle (i.e., the trajectory traced out by the reflection coefficient $S_{11}(\Delta)$ in the complex plane as a function of detuning $\Delta$) and thereby lead to systematic errors in the extracted damping rates if not properly taken into account.  Furthermore, the qualitative features of this deformation depend on the underlying source of frequency fluctuations.  Here we report the first observation of this phenomenon both for the Gaussian-distributed frequency fluctuations due to gate and flux noise, and for the chi-square-distributed frequency fluctuations due to quantum fluctuations in the cavity field.  

By using our model for the deformed resonance circles as a fitting function for experimental data, we are able to extract both the true damping rates of the cCPT and the standard deviation of frequency fluctuations $\sigma_{\omega_{0}}$ as a function of gate and flux.  We find excellent agreement between $\sigma_{\omega_{0}}$ and our model for its dependence on the gate and flux parameters, from which we extract the standard deviation of the underlying charge and flux noise.  To corroborate these results we directly measure the power spectral density of frequency fluctuations at both a charge-sensitive/flux-insensitive point and a charge-insensitive/flux-sensitive point, from which we determine the underlying power spectral densities of charge and flux fluctuations.  These power spectra follow distinct $f^{-\alpha}$ power laws where $\alpha$ is order unity, a common observation in solid state systems \cite{Paladino2014}, likely due to fluctuating two-level systems in the case of charge noise \cite{Astafiev2006,Kafanov2008} and unpaired surface spins in the case of flux noise \cite{Sendelbach2008,Kumar2016}.  We find that the scales of these power laws are in order-of-magnitude agreement with our results from fitting to the deformed resonance circles, and we discuss several limitations of comparing the two measurement schemes.

We perform most of our characterization measurements at the sub-photon level, since several of the theoretical models we use are only valid in this regime.  It is therefore essential for us to determine the number of intracavity photons in-situ, which we do by measuring the power-dependent shift in the resonant frequency induced by the Kerr nonlinearity $K$ \cite{Yurke2006, Krantz2013, thesis_krantz}.  This measurement both enables us to refer our input and output powers to the plane of the sample, and corroborates our model for how $K$ varies with gate and flux.

The paper is organized as follows.  In Section \ref{sec:cCPT} we describe the experimental realization of the cCPT and present a brief derivation of its Hamiltonian.  In Section \ref{sec:reflection_measurements} we describe how we perform and calibrate our reflection measurements, which are the primary means by which we characterize the cCPT, and discuss how these measurements are affected by frequency fluctuations.  In Section \ref{sec:tunable_resonant_frequency} we measure the tunable resonant frequency and compare our results with theory.  In Section \ref{sec:damping_rates_and_deformed_resonance_circles} we study the deformation of our resonance circles induced by frequency fluctuations and determine the internal and external damping rates of the cCPT by accounting for this effect.  In Section \ref{sec:power_spectra_of_frequency_fluctuations} we measure the power spectral densities of charge and flux fluctuations, which corroborate our results from Section \ref{sec:damping_rates_and_deformed_resonance_circles}.  Finally, in Section \ref{sec:kerr_shift} we measure the power-dependent shift in resonant frequency due to the Kerr nonlinearity, which validates several methods used in the preceding measurements.

\section{The Cavity-embedded Cooper Pair Transistor} \label{sec:cCPT}

\subsection{Experimental Design}

\begin{figure}[!t]
\includegraphics{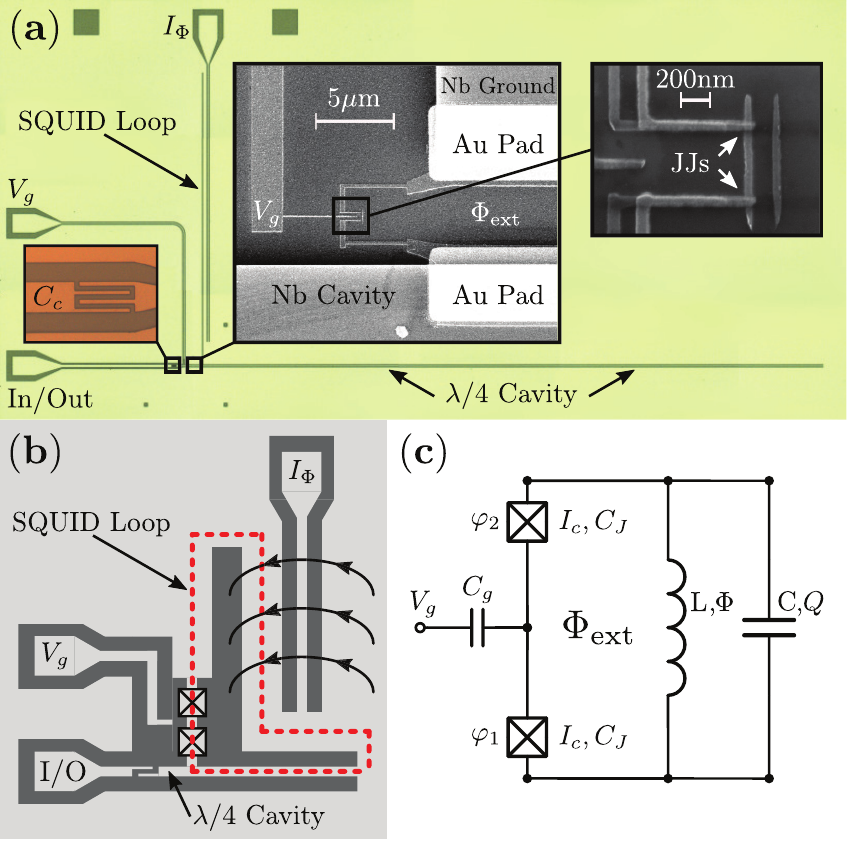}
  \caption{(a) Sample images of the cCPT. (b) An illustration of the chip layout of the cCPT; the dashed red line depicts the SQUID loop, and the black arcs depict the magnetic field lines generated by the current $I_{\Phi}$.  (c) An equivalent-circuit schematic of the closed-system cCPT.  The $\lambda/4$ cavity behaves as a parallel LC circuit when operated near its fundamental frequency $\omega_{\lambda/4}=1/\sqrt{LC}$ \cite{text_pozar}.}
  \label{fig:sample_and_schematic}
\end{figure}

The cCPT consists of a quarter-wavelength ($\lambda/4$) coplanar waveguide cavity with a Cooper pair transistor connected between its voltage antinode and the ground plane, as shown in Figure \ref{fig:sample_and_schematic}. 
The CPT is made up of two Josephson junctions (JJs) separated by an island, which can be gated with a voltage $V_{g}$ via the capacitance $C_{g}$.  Here we treat the JJs as identical, since their asymmetry is sufficiently small that it is not necessary to account for our experimental observations.  The cavity and CPT form a closed loop, which is superconducting when cooled to base temperature in a dilution refrigerator.  This SQUID loop is L-shaped, as shown in Figure \ref{fig:sample_and_schematic}(b).  The vertical segment of the loop runs parallel to a transmission line carrying the current $I_{\Phi}$, which threads flux through the loop, while the horizontal segment runs parallel to the cavity.  This design minimizes the coupling between the cavity and the transmission line carrying $I_{\Phi}$, such that the effect on the intrinsic cavity damping rate is negligible.  The cavity is driven and measured via an external transmission line coupled to its voltage antinode via the interdigitated capacitor $C_{c}$.  The cavity, input/output line, gate bias line, and flux bias line are all designed to have characteristic impedances of $Z_{0} = 50 \Omega$.  The cavity has a length $\ell = 5135$ $\mu$m and bare resonant frequency $\omega_{\lambda/4} = 2\pi\times 5.757$ GHz, which includes the slight renormalization due to the coupling capacitance $C_{c}$ (see Appendix \ref{sec:classical_circuit_model}).  

To fabricate the sample, a $100$ nm layer of Nb was first sputtered onto an intrinsic high-resistivity silicon substrate.  The on-chip transmission lines (cavity, input/output, gate/flux bias lines, and interdigitated coupling capacitor) were then patterned using photolithography, and the Nb in the negative space was removed by reactive-ion etching.  Next, the oxide layer on the Nb was removed via ion milling and $10$ nm of Au was deposited for contact pads.  Finally, the JJs were patterned using electron-beam lithography to have a cross-sectional area of roughly $50\times 50$ nm$^{2}$, and Al was deposited using a double layer shadow evaporation with an oxidation step between the layers to form the insulating barrier.  The lower layer of Al forming the island, deposited using a cryogenically-cooled stage, is $9$ nm thick so as to increase the superconducting gap energy \cite{Yamamoto2006,Court2007} and thereby suppress quasiparticle poisoning \cite{Aumentado2004}.  The upper layer of Al connecting to the Au contact pads is $65$ nm thick.  The superconducting phase around the SQUID loop remains coherent across the Au contact pads due to the proximity effect.  The fabrication techniques were similar to those described in Ref. \cite{Chen2014}, which provides further detail.

\begin{figure}[!t]
\includegraphics{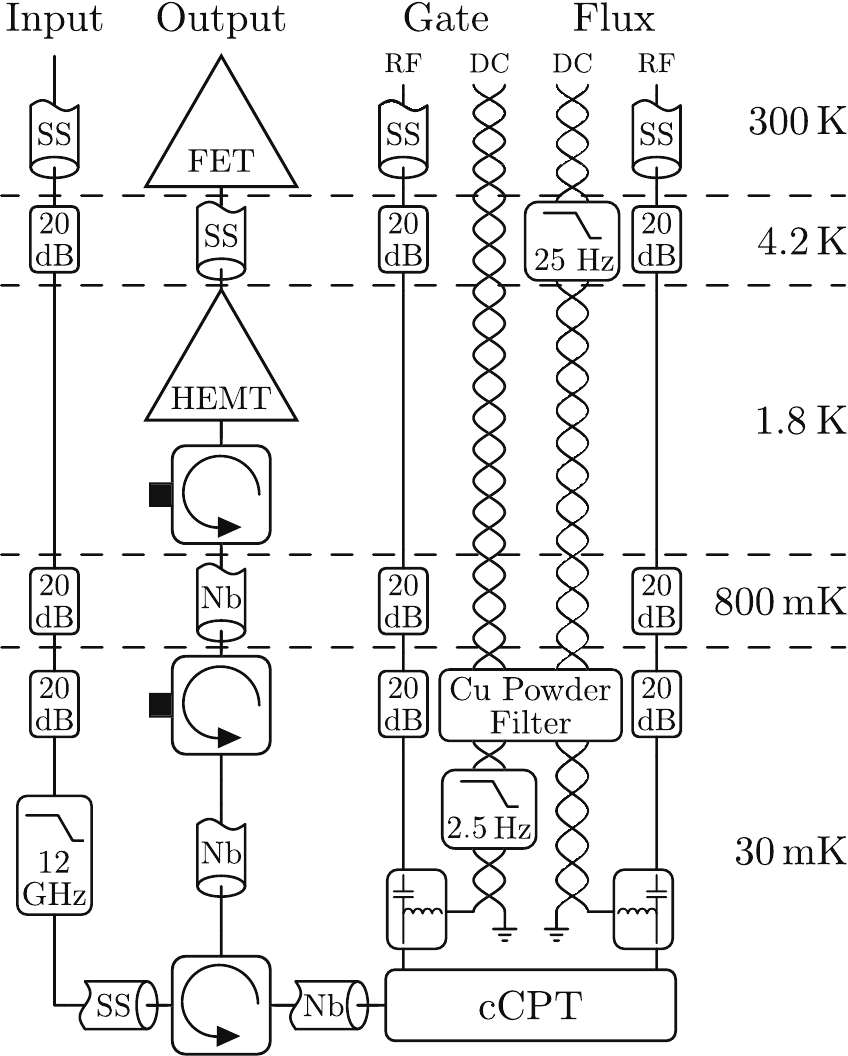}
  \caption{Schematic of the circuitry used to measure the cCPT.}
  \label{fig:measurement_setup}
\end{figure}

The sample is housed in a dilution refrigerator with a base temperature of $T\approx 30$ mK, and the sample box itself is mounted within a magnetic shield made of Cryoperm 10.  We measure the sample using the circuitry depicted in Figure \ref{fig:measurement_setup}.  On the input line, attenuators are distributed such that the input noise at $30$ mK is thermalized, and a cryogenic filter is used to suppress high frequency noise.  On the output line, a circulator is used to separate the outgoing signal, which passes through two isolators before being amplified by a cryogenic HEMT amplifier and then a room temperature FET amplifier.  Stainless steel coaxial cables are used to carry the input signal down to the circulator, giving rise to additional input attenuation, whereas niobium coaxial cables are used between the sample and HEMT to minimize attenuation of the output signal.  The gate and flux DC biases are carried by twisted pairs, which are filtered both at room temperature and at cryogenic temperatures using a combination of RC filters to suppress $60$ Hz noise, and a copper powder filter to suppress radio frequency (RF) noise \cite{Fukushima1997,thesis_masluk}.  The different size and location of the RC filter on the flux line relative to the gate line is due to the resistive heating associated with driving a steady current $I_{\Phi}$ through the flux line, compared to the negligible heating associated with maintaining a voltage $V_{g}$ across the capacitance $C_{g}$.  The gate and flux DC lines are combined with their RF counterparts via bias tees.  These RF lines can be used to apply parametric drives to the cCPT, and they are designed similarly to the input line.

\subsection{Hamiltonian}\label{sec:cCPT_Hamiltonian}

We present a simple derivation of the cCPT Hamiltonian based on the schematic depiction in Fig. \ref{fig:sample_and_schematic}(c).  For a rigorous derivation from first principles see our companion paper \cite{Kanhirathingal2020}.  The total Hamiltonian is the sum of contributions from the cavity and the CPT, such that
\begin{equation}
H = H_{\mrm{Cavity}} + H_{\mrm{CPT}}.
\end{equation}
Near its fundamental frequency, the cavity can be modeled as a lumped LC circuit with charge coordinate $Q$ and flux coordinate $\Phi$, such that
\begin{equation}
H_{\mrm{Cavity}} = \frac{Q^{2}}{2C} + \frac{\Phi^{2}}{2L}
\end{equation}
where the effective inductance and capacitance can be expressed as \cite{text_pozar}
\begin{equation}
L = \frac{4 Z_{0}}{\pi\omega_{\lambda/4}} \quad \quad \quad C = \frac{\pi}{4Z_{0}\omega_{\lambda/4}}
\end{equation}
in terms of the bare resonant frequency $\omega_{\lambda/4}$ and the characteristic impedance $Z_{0}$ of the transmission line.  The CPT Hamiltonian consists of two parts: the electrostatic energy associated with the island having $N$ excess Cooper pairs while being gated by $n_{g} = C_{g}V_{g}/e$ electrons, and the energy associated with Cooper pairs tunneling on to and off of the island.  The scale of the former is the charging energy $E_{C} = e^{2}/2C_{\Sigma}$, where $C_{\Sigma} = C_{g} + 2C_{J}$ is the total capacitance of the island, while the scale of the latter is the Josephson energy $E_{J} = I_{c}\Phi_{0}/2\pi$, where $I_{c}$ is the critical current of each Josephson junction (JJ) and $\Phi_{0}$ is the magnetic flux quantum.  The CPT Hamiltonian therefore takes the form \cite{thesis_cottet}
\begin{multline}\label{eq:CPT_hamiltonian}
H_{\mrm{CPT}} = 4E_{C}\sum\limits_{N\in\mathbb{Z}}\left(N - \frac{n_{g}}{2}\right)^{2}\ketbra{N}{N} \\
 - E_{J}\cos(\phi/2)\sum\limits_{N\in\mathbb{Z}}\Bigl(\ketbra{N+1}{N}+\ketbra{N}{N+1}\Bigr)
\end{multline}
where $\phi = \varphi_{1}+\varphi_{2}$ is the total phase across the JJs.  Since the flux through a closed superconducting loop must be an integer multiple of $\Phi_{0}$, we also have the constraint
\begin{equation}\label{eq:SQUID_loop_superconducting_phase}
\frac{\Phi_{0}}{2\pi}\left(\varphi_{1}+\varphi_{2}\right) - \Phi - \Phi_{\ext} = m\Phi_{0}
\end{equation}
for some constant integer $m$, assuming a negligible self-inductance of the SQUID loop (we analyze the validity of this assumption in Appendix \ref{sec:SQUID_self_inductance}).  Absorbing the constant multiple of $\Phi_{0}$ into $\Phi_{\ext}$, the total phase across the JJs can be expressed as
\begin{equation}
\phi = \frac{2\pi}{\Phi_{0}}\left(\Phi + \Phi_{\ext}\right).
\end{equation}
Thus, the cavity and CPT couple to one another via their shared SQUID loop.

We designed our system such that the bare resonant frequency of the cavity $\omega_{\lambda/4} = 1/\sqrt{LC}$ is much smaller than the frequency required to drive the CPT to its first excited state (on the order of $E_{J}/\hbar$ and $E_{C}/\hbar$), provided we operate sufficiently far from the points at which the ground and first excited states are degenerate.  In this case the CPT remains in its ground state when the cavity is driven near its fundamental frequency, and the ground state energy $E_{\mathrm{CPT}}(n_{g}, \Phi_{\ext}, \Phi)$ contributes to the effective potential of the cavity.  We analyze the cCPT Hamiltonian by Taylor expanding this ground state energy in powers of the cavity flux coordinate $\Phi$ according to
\begin{equation}\label{eq:cCPT_expansion_full}
H = \frac{Q^{2}}{2C} + \frac{\Phi^{2}}{2L}
+ \sum_{k=0}^{\infty}\frac{1}{k!}\partial^{k}_{\phi}E_{CPT}(n_{g},\Phi_{\ext})\left(\frac{2\pi\Phi}{\Phi_{0}}\right)^{k}
\end{equation}
where we have introduced the shorthand notation
\begin{equation}
\partial^{k}_{\phi}E_{CPT}(n_{g},\Phi_{\ext}) = \frac{\partial^{k} E_{\mathrm{CPT}}(n_{g}, \phi)}{\partial \phi^{k}}\Biggr|_{\phi = 2\pi\Phi_{\ext}/\Phi_{0}}
\end{equation}
for convenience.  The $k=0$ and $k=1$ terms in this expansion can be dropped, since the former yields a constant offset to the Hamiltonian and the latter yields a negligibly small shift in the equilibrium flux coordinate of the cavity.  We next combine the terms proportional to $\Phi^{2}$ to express the Hamiltonian as
\begin{equation}
H = \frac{Q^{2}}{2C} + \frac{\Phi^{2}}{2L_{\mathrm{tot}}}
+ \sum_{k=3}^{\infty}\frac{1}{k!}\partial^{k}_{\phi}E_{CPT}(n_{g},\Phi_{\ext})\left(\frac{2\pi\Phi}{\Phi_{0}}\right)^{k}
\end{equation}
where the total inductance is given by
\begin{equation}
\frac{1}{L_{\mathrm{tot}}} = \frac{1}{L} + \left(\frac{2\pi}{\Phi_{0}}\right)^{2}\partial^{2}_{\phi}E_{CPT}(n_{g},\Phi_{\ext}).
\end{equation}
This is simply the parallel combination of the bare inductance $L$ with the tunable Josephson inductance \cite{thesis_sillanpaa}
\begin{equation}
L_{J}(n_{g},\Phi_{\mathrm{ext}}) = \left(\frac{\Phi_{0}}{2\pi}\right)^{2}\Bigl[\partial^{2}_{\phi}E_{CPT}(n_{g},\Phi_{\ext})\Bigr]^{-1}
\end{equation}
where $L_{J}\gg L$ over the full range of $n_{g}$ and $\Phi_{\mathrm{ext}}$.  

We can now proceed to quantize this system by imposing the canonical commutation relation $[\Phi, Q] = i\hbar$ and introducing the cavity mode operator
\begin{equation}\label{eq:cavity_mode_operator}
a = \frac{1}{\sqrt{2\hbar Z}}\left(\Phi+iZQ\right)
\end{equation}
where $Z = \sqrt{L_{\mathrm{tot}}/C} \approx 4Z_{0}/\pi$ is the mode impedance, whose deviation from this value due to $L_{J}$ is negligible.  The Hamiltonian can then be expressed as
\begin{equation}
H = \hbar\omega_{0} a^{\dagger}a + \sum_{k=3}^{\infty}\frac{\phi_{\mathrm{zp}}^{k}}{k!}\partial^{k}_{\phi}E_{CPT}(n_{g},\Phi_{\ext})\left(a+a^{\dagger}\right)^{k}
\end{equation}
where $\omega_{0} = 1/\sqrt{L_{\mathrm{tot}}C}$ and we have introduced the dimensionless quantity
\begin{equation}\label{eq:phizp_val}
\phi_{\mathrm{zp}} = \frac{2\pi}{\Phi_{0}}\sqrt{\frac{\hbar Z}{2}} \approx 0.176
\end{equation}
which is the scale of zero point fluctuations in the total phase across the JJs.  To leading order in $L/L_{J}$, the resonant frequency can be expressed as 
\begin{equation}\label{eq:tunable_resonant_frequency}
\omega_{0}(n_{g}, \Phi_{\mathrm{ext}}) = \omega_{\lambda/4} + \frac{\phi_{\mathrm{zp}}^{2}}{\hbar}\partial^{2}_{\phi}E_{CPT}(n_{g},\Phi_{\ext})
\end{equation}
which can be tuned by both $n_{g}$ and $\Phi_{\mathrm{ext}}$.

To analyze the nonlinear terms in the Hamiltonian we perform a rotating wave approximation, keeping only those terms with like powers of $a$'s and $a^{\dagger}$'s that will be stationary in a frame rotating near the resonant frequency of the cCPT.  Doing so, we find
\begin{equation}
H = \hbar\omega_{0} a^{\dagger}a + \sum_{k=2}^{\infty}\frac{\phi_{\mathrm{zp}}^{2k}}{(k!)^{2}}\partial^{2k}_{\phi}E_{CPT}(n_{g},\Phi_{\ext})a^{\dagger k} a^{k}
\end{equation}
where we have ignored the negligible renormalization of the lower order terms in the expansion due to the higher order terms.   The leading-order nonlinear term is that due to the Kerr nonlinearity 
\begin{equation}\label{eq:tunable_kerr_nonlinearity}
K(n_{g},\Phi_{\mathrm{ext}}) = \frac{\phi_{\mathrm{zp}}^{4}}{2\hbar}\partial^{4}_{\phi}E_{CPT}(n_{g},\Phi_{\ext})
\end{equation}
defined relative to the standard form of $H_{\mathrm{Kerr}} = \hbar K a^{\dagger 2}a^{2}/2$.  

In the present work we operate at sufficiently small photon numbers that we can truncate the Hamiltonian at this term, such that our effective Hamiltonian takes the form
\begin{equation}
H = \hbar\omega_{0} a^{\dagger}a + \frac{1}{2}\hbar Ka^{\dagger 2}a^{2}
\end{equation}
where $\omega_{0}$ and $K$ are both tunable and given by Eqs. \eqref{eq:tunable_resonant_frequency} and \eqref{eq:tunable_kerr_nonlinearity}, respectively.  To compare our experimental results with theory, we evaluate the ground state energy $E_{\mathrm{CPT}}$ and its derivatives numerically by including only the five lowest energy charge states in the expansion of the CPT Hamiltonian given by Eq. \eqref{eq:CPT_hamiltonian}.  This approximation is very accurate in our case, since we are operating well into the Cooper pair box regime where $E_{C} \gtrsim E_{J}$ \cite{Koch2007}.  

It is worth noting that parametric pumping of the flux line near $2\omega_{0}$ yields an additional term in the Hamiltonian that will be close to stationary in a frame rotating near the resonant frequency of the cCPT.  We have neglected this term in the above analysis since we do not incorporate parametric pumping in any of the measurements presented in this work.  For completeness, however, we have derived this additional term in Appendix \ref{sec:cCPT_hamiltonian_parametric_pumping}.  

\section{Reflection Measurements}\label{sec:reflection_measurements}

We characterize the cCPT using a vector network analyzer (VNA) to measure the reflection coefficient $S_{11}$ at the plane of the sample by means of the transmission coefficient $S_{21}^{\mathrm{VNA}}$ from the input to the output port.  The two are related according to
\begin{equation}
S_{21}^{\mathrm{VNA}}(\omega) = \frac{G(\omega)}{\eta_{\mathrm{in}}}e^{i\theta(\omega)}S_{11}(\Delta)
\end{equation}
where $\eta_{\mathrm{in}}$ is the input attenuation, $G(\omega)$ is the gain of the amplifier chain, $\theta(\omega)$ is an overall phase shift, and $\Delta = \omega - \omega_{0}$ is the detuning from resonance.  We treat the input attenuation as constant since it should not vary significantly over the tuning range of the cCPT ($\approx 140$ MHz).  As shown in Appendices \ref{sec:classical_circuit_model} and \ref{sec:linear_cavity_response}, the reflection coefficient of the cCPT takes the form
\begin{equation}\label{eq:s11_linear}
S_{11}(\Delta) = \frac{\Delta - i(\kappa_{\mathrm{int}} - \kappa_{\mathrm{ext}})/2}{\Delta - i(\kappa_{\mathrm{int}} + \kappa_{\mathrm{ext}})/2}
\end{equation}
when operated in the linear response regime, where $\kappa_{\mathrm{int}}$ and $\kappa_{\mathrm{ext}}$ are the damping rates associated with internal loss and coupling to the external transmission line, respectively.  We find the prefactor $G(\omega)e^{i\theta(\omega)}/\eta_{\mathrm{in}}$ by measuring $S_{21}^{\mathrm{VNA}}$ far off resonance, since $S_{11}(\Delta)\approx 1$ for $|\Delta| \gg \kappa_{\mathrm{int}}+\kappa_{\mathrm{ext}}$, which enables us to determine $S_{11}$ from measurements of $S_{21}^{\mathrm{VNA}}$.  Usually the value of this prefactor at resonance is inferred from an off-resonant measurement, but in our case we can perform this calibration at any frequency by detuning $\omega_{0}$ itself.  We can then fit our theoretical model for $S_{11}(\Delta)$ to our measurements to extract the physical parameters characterizing the cCPT.  To do so accurately, however, we must properly account for the effect of frequency fluctuations on our model for $S_{11}(\Delta)$.

For our analysis it is convenient to rewrite the linear reflection coefficient from Eq. \eqref{eq:s11_linear} as
\begin{equation}
S_{11}(\Delta) = \frac{\kappa_{\mathrm{int}}}{\kappa_{\mathrm{tot}}} - \frac{\kappa_{\mathrm{ext}}}{\kappa_{\mathrm{tot}}}e^{-2i\arctan\left(2\Delta/\kappa_{\mrm{tot}}\right)}
\end{equation}
where $\kappa_{\mathrm{tot}} = \kappa_{\mathrm{int}} + \kappa_{\mathrm{ext}}$ is the total damping rate of the cCPT.  From this expression it is clear that $S_{11}$ traces out a circle in the complex plane as a function of detuning $\Delta$, such that $\kappa_{\mathrm{tot}}$ can be determined from the rate of traversal, $\kappa_{\mathrm{ext}}$ from the radius, and $\omega_{0}$ from the intercept with the real axis (or, equivalently, from the minimum of $|S_{11}|$) \cite{Probst2015}.  However, this treatment is no longer accurate in the presence of frequency fluctuations comparable in scale to the cavity linewidth $\kappa_{\mathrm{tot}}$, as was recently shown \cite{Brock2020}.  Following the analysis of Ref.~\cite{Brock2020}, we model the effect of frequency fluctuations by letting $\omega_{0}\rightarrow \omega_{0}+\delta\omega_{0}$ where $\delta\omega_{0}$ is treated as a random variable.  The average reflection coefficient measured by the VNA will then be the convolution
\begin{equation}\label{eq:avg_s11_convolution}
\overline{S_{11}}(\Delta) = \int\limits_{-\infty}^{\infty}S_{11}(\Delta-\Omega)P(\Omega)d\Omega
\end{equation}
where $P(\Omega)$ is the probability density function associated with drawing the value $\Omega$ from the random variable $\delta\omega_{0}$.  This convolution causes a deformation of the resonance circle, yielding an apparent decrease in the radius and increase in the linewidth that result in systematic errors if not taken into account.  Furthermore, different probability distributions give rise to qualitatively different deformations, which provide evidence of the underlying source of the frequency fluctuations.  Here we must consider two sources of frequency fluctuations: those induced by fluctuations in the tuning parameters $n_{g}$ and $\Phi_{\mathrm{ext}}$, and those induced by quantum fluctuations in the cavity field via the Kerr nonlinearity.  

To linear order in the fluctuations $\delta n_{g}$ and $\delta \Phi_{\mathrm{ext}}$, the frequency fluctuations they induce take the form
\begin{equation}\label{eq:tuning_fluctuations_frequency_coupling}
\delta\omega_{0} = \frac{\partial\omega_{0}}{\partial n_{g}}\delta n_{g} + \frac{\partial\omega_{0}}{\partial \Phi_{\ext}}\delta \Phi_{\mathrm{ext}}.
\end{equation}
Assuming fluctuations in the gate and flux are independent and Gaussian-distributed with mean zero and variances $\sigma_{n_{g}}^{2}$ and $\sigma_{\Phi_{\mathrm{ext}}}^{2}$, respectively, then these frequency fluctuations will be Gaussian-distributed with mean zero and variance
\begin{equation}\label{eq:tuning_fluctuations_frequency_variance}
\sigma_{\omega_{0}}^{2} = \abs{\frac{\partial\omega_{0}}{\partial n_{g}}}^{2} \sigma_{n_{g}}^{2} + \abs{\frac{\partial\omega_{0}}{\partial \Phi_{\ext}}}^{2} \sigma_{\Phi_{\mathrm{ext}}}^{2}.
\end{equation}
In this case the average reflection coefficient takes the form \cite{Brock2020}
\begin{equation}\label{eq:avg_s11_gaussian}
\overline{S_{11}}(\Delta) = 
1 - \sqrt{\frac{\pi}{2}}\frac{\kappa_{\mathrm{ext}}}{\sigma_{\omega_{0}}} w\left(\frac{i\kappa_{\mathrm{tot}}-2\Delta}{2\sqrt{2}\sigma_{\omega_{0}}}\right)
\end{equation}
where $w(z) = e^{-z^{2}}\mathrm{erfc}(-iz)$ is the Faddeeva function~\cite{text_NIST_handbook}.  

On the other hand, the frequency fluctuations induced by quantum fluctuations in the cavity field via the Kerr nonlinearity take the form
\begin{equation}
\delta\omega_{0} = \frac{K}{4}(X_{1}^{2} + X_{2}^{2})
\end{equation}
where $X_{1} = a^{\dagger}+a$ and $X_{2} = i(a^{\dagger}-a)$ are the quadrature operators.  Assuming a coherent steady state of the cavity field, fluctuations $\delta X_{1,2} = X_{1,2} - \langle X_{1,2} \rangle$ in these operators will be independent and Gaussian, each with zero mean and unit variance.  In the case of small cavity occupation $n \lesssim 1/4$, the resulting frequency fluctuations follow a chi-square distribution with two degrees of freedom to good approximation, such that the average reflection coefficient takes the form \cite{Brock2020}
\begin{equation}\label{eq:avg_s11_kerr}
\overline{S_{11}}(\Delta) = 1 - 2i\frac{\kappa_{\mathrm{ext}}}{K}e^{(i\kappa_{\mathrm{tot}}-2\Delta)/K}\Gamma\Bigl(0,\frac{i\kappa_{\mathrm{tot}}-2\Delta}{K}\Bigr)
\end{equation}
where $\Gamma(a, z) = \int_{z}^{\infty}t^{a-1}e^{-t}dt$ is the incomplete gamma function \cite{text_NIST_handbook}.  

In general, we must consider the combined effect of these two sources of fluctuations, which leads to a complicated probability distribution that requires the convolution in Eq. \eqref{eq:avg_s11_convolution} to be evaluated numerically.  Of particular importance to us, however, is when the frequency fluctuations are dominated by fluctuations in the gate and flux such that $\sigma_{\omega_{0}} \gtrsim K/2$.  In this case $\delta\omega_{0}$ will be Gaussian-distributed to good approximation, its variance will be renormalized by the Kerr nonlinearity according to \cite{Brock2020}
\begin{equation}\label{eq:renormalized_variance}
\sigma_{\omega_{0}}^{2} = \abs{\frac{\partial\omega_{0}}{\partial n_{g}}}^{2} \sigma_{n_{g}}^{2} + \abs{\frac{\partial\omega_{0}}{\partial \Phi_{\ext}}}^{2} \sigma_{\Phi_{\mathrm{ext}}}^{2} + \frac{K^{2}}{4}
\end{equation}
and the average reflection coefficient will be given by Eq. \eqref{eq:avg_s11_gaussian}.  By fitting these models for the average reflection coefficient to our measurements, we can extract the true damping rates of the cCPT in the presence of frequency fluctuations.  

In addition to the deformation induced by frequency fluctuations, we have observed that the trajectories traced out by $S_{11}$ in our experiments are rotated about the off-resonant point $S_{11} = 1$.  This is a sign of impedance mismatching at the sample input, likely due to the self-inductance of wire bonds, which also causes a dilation of the resonance circle \cite{Megrant2012,Khalil2012}.  In our case, the angle of rotation remains less than $0.1$ radians in magnitude over the full tuning range of the cCPT, which leads to a systematic error in our extracted damping rates of less than $0.5\%$.  Since this is generally smaller than their confidence intervals we can safely ignore this effect.  It is also worth noting that this rotation angle can be extracted by finding the tangent to the trajectory at $S_{11}=1$, which is independent of all other fitting parameters.  Thus, although we account for these rotation angles in our data analysis, for the sake of clarity we do not report them.

In sections \ref{sec:tunable_resonant_frequency} and \ref{sec:damping_rates_and_deformed_resonance_circles} we characterize the cCPT by measuring the reflection coefficient using an input VNA power of $P_{\mathrm{VNA}} = -65$ dBm, which yields a maximum cavity occupation of $n_{\mathrm{max}} = 0.28$ photons over the full tuning range of the cCPT.  This is both well within the linear-response regime ($n_{\mathrm{max}}|K|/\kappa_{\mathrm{tot}}\ll 1$) and sufficiently small that Eq. \eqref{eq:avg_s11_kerr} is appropriate for modeling the effect of quantum fluctuations on the average reflection coefficient.  Calibration of the intra-cavity photon number is performed in Section \ref{sec:kerr_shift} by measuring the power-dependent shift of the resonant frequency induced by the Kerr nonlinearity \cite{Yurke2006,Krantz2013}, which enables us to refer our input power to the plane of the sample.

\section{Tunable Resonant Frequency}\label{sec:tunable_resonant_frequency}
\begin{figure}[!t]
  \includegraphics{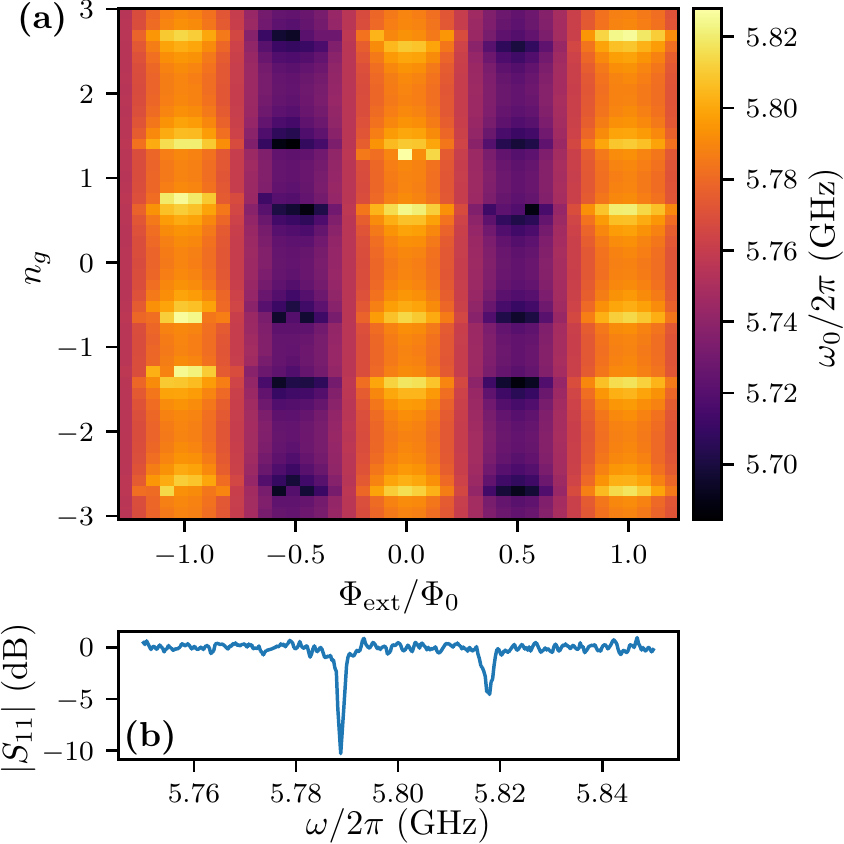}
  \caption{(a) Resonant frequency $\omega_{0}(n_{g},\Phi_{\mathrm{ext}})$ measured over multiple periods of $n_{g}$ and $\Phi_{\mathrm{ext}}$.  (b) VNA trace of $|S_{11}|$ at the quasiparticle poisoning threshold $(n_{g},\Phi_{\mathrm{ext}}) = (0.7,0)$, showing both the even and odd parity resonances.}
  \label{fig:multiperiod_f0}
\end{figure}

We first measure the resonant frequency $\omega_{0}(n_{g},\Phi_{\ext})$ over multiple periods of the gate and flux, the results of which are presented in Figure \ref{fig:multiperiod_f0}(a).  
From its periodicity we extract the gate capacitance $C_{g} = 6.3$ aF and the mutual inductance $M_{\Phi} = 42$ pH between the transmission line carrying the current $I_{\Phi}$ and the SQUID loop of the cCPT.  These enable us to convert $V_{g}$ and $I_{\Phi}$ into units of $n_{g}$ and $\Phi_{\mathrm{ext}}$, respectively.

A prominent feature of $\omega_{0}(n_{g},\Phi_{\mathrm{ext}})$ is the presence of sudden jumps when $\abs{(n_{g}-1)\mod 2}\approx 0.3$ due to quasiparticle poisoning \cite{Naaman2006}.  Near this point, the decrease in energy obtained from the transition $n_{g}\rightarrow n_{g}+1$ is comparable to the energy required for quasiparticles to tunnel onto the island, so quasiparticles can tunnel back and forth \cite{Aumentado2004}.  The latter energy scale is the difference between the superconducting gaps of the island and the leads, $\delta\Delta = \Delta_{i} - \Delta_{l}$, which arises in our case due to the island's thickness of $9$ nm relative to the leads' thickness of $65$ nm \cite{Yamamoto2006,Court2007}.  Closer to $n_{g} \equiv 0 \mod 2$ the even parity ground state is energetically favorable, whereas closer to $n_{g} \equiv 1 \mod 2$ the odd parity ground state is energetically favorable.  The two states become equiprobable at the critical gate charge $n_{g}^{c} \approx 0.7$ such that
\begin{equation}\label{eq:critical_quasiparticle_poisoning}
E_{\mrm{CPT}}(n_{g}^{c},\Phi_{\ext}) - E_{\mrm{CPT}}(n_{g}^{c}+1,\Phi_{\mathrm{ext}}) = \delta\Delta.
\end{equation}
We note that for $E_{C}>E_{J}$ the left hand side of the above equation does not vary appreciably with $\Phi_{\mathrm{ext}}$, which is why the threshold $n_{g}^{c}$ does not vary appreciably with $\Phi_{\mathrm{ext}}$ either.  Near this threshold, random switching occurs between the even and odd parity states.  As we show in Appendix \ref{sec:power_spectrum_of_quasiparticle_poisoning}, the power spectral density of these switching events follows a Lorentzian with corner frequency $830$ Hz, consistent with other reports in the literature \cite{Sun2012, Riste2013}.  When measured with a VNA, this switching manifests itself as two visible resonances as shown in Figure \ref{fig:multiperiod_f0}(b).  The blockiness of Fig. \ref{fig:multiperiod_f0}(a) around the transition point $n_{g}\approx 0.7$ is the result of identifying only one of these two resonant frequencies.  

\begin{figure}[!t]
  \includegraphics{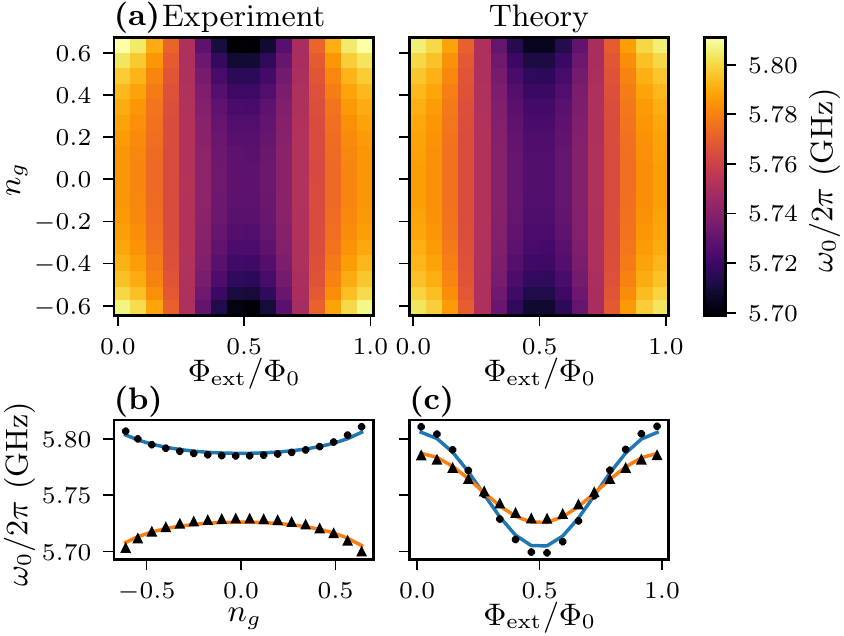}
  \caption{(a) Measured resonant frequency $\omega_{0}(n_{g},\Phi_{\mathrm{ext}})$ and the best fit to Eq. \eqref{eq:tunable_resonant_frequency}.  Cross-sections of (a) are plotted in (b) and (c).  In (b), circles correspond to $\Phi_{\mathrm{ext}} = 0$ and triangles correspond to $\Phi_{\mathrm{ext}} = \Phi_{0}/2$.  In (c), circles correspond to $n_{g} = 0.64$ and triangles correspond to $n_{g} = 0$.  Solid lines are the corresponding cross-sections of the best fit.}
  \label{fig:tunable_f0_theory_exp}
\end{figure}

For our remaining measurements we restrict ourselves to the region $-0.65 \leq n_{g} \leq 0.65$ to avoid the effects of quasiparticle poisoning.  In this region we can fit Eq. \eqref{eq:tunable_resonant_frequency} to our measurements of $\omega_{0}(n_{g},\Phi_{\mathrm{ext}})$ to extract $E_{J}$ and $E_{C}$.  As shown in Figure \ref{fig:tunable_f0_theory_exp}, we find excellent agreement between theory and experiment, and obtain the best fit parameters
\begin{equation}\label{eq:Ej_and_Ec}
\begin{split}
E_{J}/h &= 14.80 \pm 0.04 \: \mathrm{GHz} \\
E_{C}/h &= 54.1 \pm 0.2 \: \mathrm{GHz}.
\end{split}
\end{equation}
This extracted value of the Josephson energy is consistent with the normal resistance of each junction, which is about $10$ k$\Omega$.  This resistance cannot be measured directly on the sample we are studying, since the source and drain of the CPT are shorted with respect to dc signals.  Rather, this estimate of the normal resistance is based on devices we made to hone the fabrication recipe, which were designed to allow such a dc measurement.  Similarly, this extracted value of the charging energy is consistent with $50\times 50$ nm$^{2}$ junctions whose oxide layer is about $1$ nm thick.  It is important to note that given these values for $E_{J}$ and $E_{C}$, the Kerr nonlinearity can now be calculated using Eq. \eqref{eq:tunable_kerr_nonlinearity}.

This agreement between our expected and extracted values of $E_{J}$ and $E_{C}$ also helps to corroborate the value of $\phi_{\mrm{zp}}\approx 0.176$ discussed in Sec. \ref{sec:cCPT_Hamiltonian}.  If the actual value of $\phi_{\mrm{zp}}$ differed considerably from $0.176$ we would have obtained best fits for $E_{J}$ and $E_{C}$ that differed considerably from their expected values as well, since the tunable resonant frequency is proportional to $\phi_{\mrm{zp}}^{2}$.  We can further corroborate this value for $\phi_{\mrm{zp}}$ by plugging our best fit values for $E_{J}$ and $E_{C}$ back into Eq. \eqref{eq:critical_quasiparticle_poisoning}, which is independent of $\phi_{\mrm{zp}}$.  Doing so, we estimate the difference between the superconducting gap energies of the island and the leads to be $\delta\Delta \approx 80$ $\mu$eV, in line with our expectations \cite{Yamamoto2006,Court2007}.  As with the normal resistance, a direct measurement of the superconducting gaps of the island and leads  cannot be made via dc transport measurements, since the source and drain of the CPT are shorted with respect to dc signals.

Qualitatively, the polynomial dependence of $\omega_{0}$ on $n_{g}$ in Fig. \ref{fig:tunable_f0_theory_exp}(b) arises from the charging energy term in the CPT Hamiltonian of Eq. \eqref{eq:CPT_hamiltonian}, while the sinusoidal dependence of $\omega_{0}$ on $\Phi_{\mathrm{ext}}$ in Fig. \ref{fig:tunable_f0_theory_exp}(c) arises from the Josephson energy term.  This behavior underscores our choice of $E_{C}$ and $E_{J}$.  The present device was designed to optimize its charge sensitivity by maximizing $\partial\omega_{0}/\partial n_{g}$, which tends to increase with the ratio $E_{C}/E_{J}$ \cite{Kanhirathingal2020,Sillanpaa2005}.  We therefore tried to maximize $E_{C}$ by minimizing the cross-sectional area of the JJs, and aimed for a Josephson energy that satisfies $\hbar\omega_{0}<E_{J}<E_{C}$ to maintain the validity of the ground-state approximation discussed in Sec. \ref{sec:cCPT_Hamiltonian}.  For other uses, such as magnetometry and parametric oscillation, a different regime of $E_{C}$ and $E_{J}$ may be optimal.

\section{Damping Rates and Deformed Resonance Circles}
\label{sec:damping_rates_and_deformed_resonance_circles}

As discussed in Section \ref{sec:reflection_measurements}, in order to extract the damping rates of the cCPT we must fit our measured reflection coefficients to an appropriate model that accounts for the effects of frequency fluctuations.  To this end we first study the trajectories traced out by our measured reflection coefficients and their deformation due to frequency fluctuations.  This will both corroborate the presence of frequency fluctuations comparable to the cavity linewidth, as well as provide evidence of the relative magnitudes of fluctuations due to gate/flux noise and those due to quantum noise.  

\begin{figure}[!t]
  \includegraphics{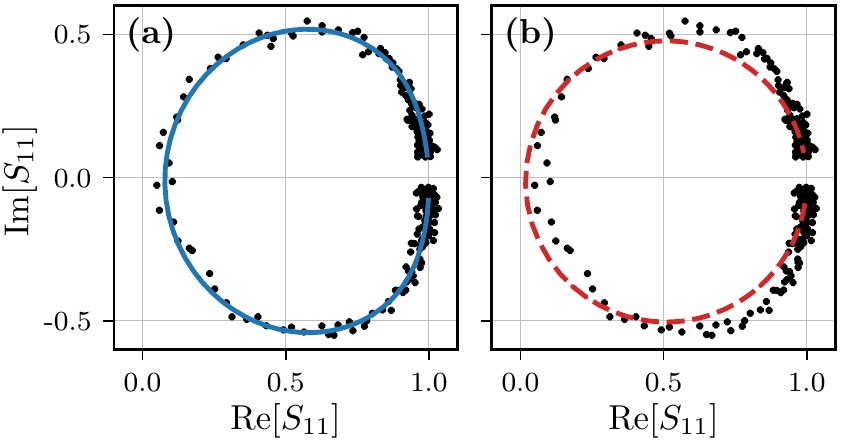}\\[6pt] 
  \begin{tabular}{c|c|c|c}  & $\kappa_{\mathrm{int}}/2\pi$ (MHz)  &  $\kappa_{\mathrm{ext}}/2\pi$ (MHz) & $\sigma_{\omega_{0}}/2\pi$ (MHz)  \\ \hline 
 (a) & $0.10\pm 0.05$ & $1.178 \pm 0.008$ & $1.04 \pm 0.02$  \\ \hline
 (b) & $1.40 \pm 0.02$ & $1.36 \pm 0.01$ &
  \end{tabular}
   \caption{Deformation of a resonance circle due to Gaussian frequency fluctuations.  The cCPT is biased to $(n_{g},\Phi_{\mathrm{ext}}) = (0.64, 0.27\Phi_{0})$, where $\omega_{0}/2\pi = 5.751$ GHz and $K/2\pi = -0.03$ MHz.  The trajectory traced out by $S_{11}(\Delta)$ is measured over a $30$ MHz span around resonance, and marked by the black dots.  The solid blue line in (a) is the fit to Eq. \eqref{eq:avg_s11_gaussian} that accounts for Gaussian frequency fluctuations, whereas the dashed red line in (b) is the fit to Eq. \eqref{eq:s11_linear} that does not account for any frequency fluctuations.  The best fit parameters are presented in the table above.}
   \label{fig:s11_gaussian_fitting_comparison}
\end{figure}

To study the effect of fluctuations in the tuning parameters $n_{g}$ and $\Phi_{\mathrm{ext}}$ on the resonance circle, we bias the cCPT to $(n_{g},\Phi_{\mathrm{ext}}) = (0.64, 0.27\Phi_{0})$.  At this point in parameter space the resonant frequency is very sensitive to both gate and flux, since both $|\partial\omega_{0}/\partial n_{g}|$ and $|\partial\omega_{0}/\partial\Phi_{\mathrm{ext}}|$ are close to their maximum values, leading to strong frequency fluctuations in accordance with Eqs. \eqref{eq:tuning_fluctuations_frequency_coupling} and \eqref{eq:tuning_fluctuations_frequency_variance}.  Furthermore, the Kerr nonlinearity $K/2\pi = -0.03$ MHz is much smaller than the cavity linewidth at this point such that quantum fluctuations will contribute negligibly to the frequency fluctuations \cite{Brock2020}.  In Fig. \ref{fig:s11_gaussian_fitting_comparison} we highlight the deformation of the resonance circle at this point by fitting our measured trajectory $S_{11}(\Delta)$ to both Eq. \eqref{eq:avg_s11_gaussian} (which accounts for Gaussian fluctuations) and Eq. \eqref{eq:s11_linear} (which does not account for any fluctuations).  It is plain to see that our measured resonance circle is deformed into an oblong shape in excellent agreement with our Gaussian model for frequency fluctuations, and which cannot be explained using the non-fluctuating model.  This both confirms the presence of strong frequency fluctuations comparable to the cavity linewidth, since $\sigma_{\omega_{0}}/\kappa_{\mathrm{tot}} \approx 0.8$ at this point, and justifies our treatment of the underlying fluctuations in the gate and flux as Gaussian-distributed random variables.  Most importantly, had we failed to account for these fluctuations we would have extracted an internal damping rate that differed from its true value by an order of magnitude.  

\begin{figure}[!t]
  \includegraphics{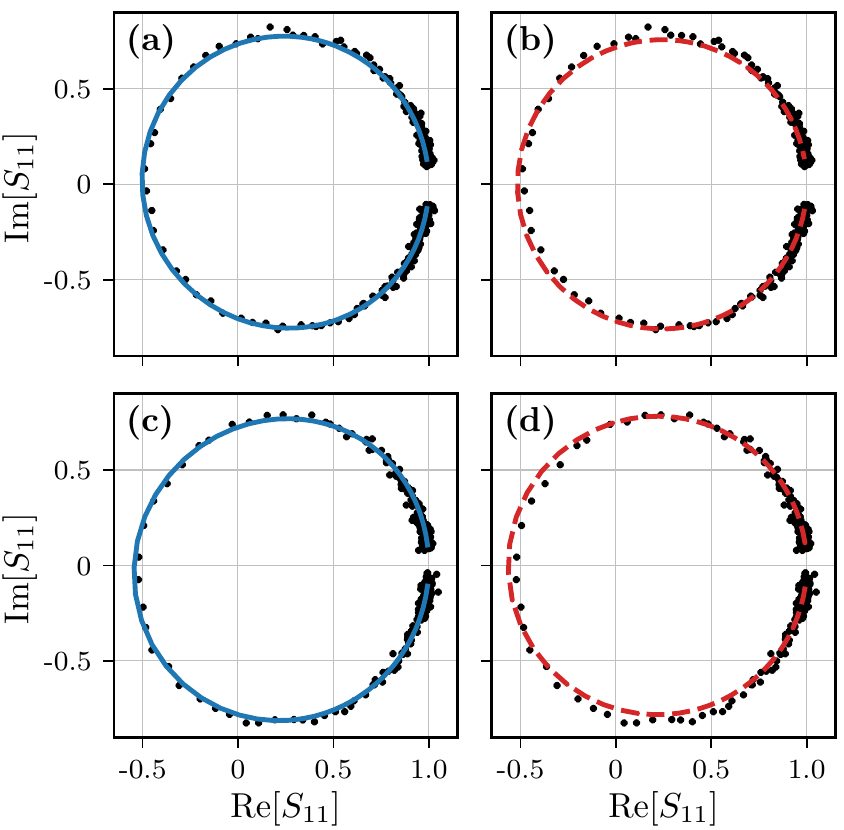}\\[6pt]
  \begin{tabular}{c|c|c}  & $\kappa_{\mathrm{int}}/2\pi$ (MHz)  &  $\kappa_{\mathrm{ext}}/2\pi$ (MHz) \\ \hline 
 (a) & $0.333 \pm 0.005$ & $1.295 \pm 0.004$ \\ \hline
 (b) & $0.420 \pm 0.006$ & $1.315 \pm 0.004$ \\ \hline 
 (c) & $0.197 \pm 0.005$ & $1.081 \pm 0.004$ \\ \hline
 (d) & $0.308 \pm 0.006$ & $1.106 \pm 0.004$
  \end{tabular}
   \caption{Deformation of two resonance circles due to quantum fluctuations.  In (a) and (b) the cCPT is biased to $(n_{g},\Phi_{\mathrm{ext}}) = (0, 0)$ where $\omega_{0}/2\pi = 5.785$ GHz and $K/2\pi = -0.46$ kHz (obtained from Eq. \eqref{eq:tunable_kerr_nonlinearity}).  In (c) and (d) it is biased to $(n_{g},\Phi_{\mathrm{ext}}) = (0, \Phi_{0}/2)$ where $\omega_{0}/2\pi =5.728$ GHz and $K/2\pi = 0.49$ MHz.  Each trajectory traced out by $S_{11}(\Delta)$ is measured over a $20$ MHz span around resonance, and marked by the black dots.  The solid blue lines in (a) and (c) are fits to Eq. \eqref{eq:avg_s11_kerr} that accounts for quantum fluctuations (using the above values for $K$), whereas the dashed red lines in (b) and (d) are fits to Eq. \eqref{eq:s11_linear} that does not account for any frequency fluctuations.  The best fit parameters are presented in the table above. }
  \label{fig:s11_chisq_fitting_comparison}
\end{figure}

To study the effect of quantum fluctuations on the resonance circle, we bias the cCPT to a point where the resonant frequency is insensitive to both gate and flux, but the Kerr nonlinearity is comparable to $\kappa_{\mathrm{tot}}$.  As it turns out there are two such points per period, $(n_{g},\Phi_{\ext}) = (0,0)$ where $K/2\pi = -0.46$ MHz and $(n_{g},\Phi_{\ext}) = (0,\Phi_{0}/2)$ where $K/2\pi = 0.49$ MHz, such that both $|\partial\omega_{0}/\partial n_{g}|$ and $|\partial\omega_{0}/\partial\Phi_{\mathrm{ext}}|$ tend toward zero at these points.  In Fig. \ref{fig:s11_chisq_fitting_comparison} we highlight the deformation of the resonance circle at these points by fitting our measured trajectories $S_{11}(\Delta)$ to both Eq. \eqref{eq:avg_s11_kerr} (which accounts for frequency fluctuations induced by quantum fluctuations via the Kerr nonlinearity) and Eq. \eqref{eq:s11_linear} (which does not account for any fluctuations).  A signature of the deformation in this case is asymmetry of the trajectory with respect to reflection of $S_{11}(\Delta)$ across the real axis \cite{Brock2020}.  This arises from the chi-square distribution of the underlying fluctuations in the cavity quadratures, which only has support on either the positive or negative reals depending on the sign of $K$.  Although it is subtle, we observe this deformation of our measured trajectories in agreement with Eq. \eqref{eq:avg_s11_kerr}; furthermore, the parity of this asymmetry depends on the sign of $K$ as expected.  In Fig. \ref{fig:s11_chisq_fitting_comparison}(b), where $K<0$, we find that our measured trajectory $S_{11}(\Delta)$ lies outside our best fit to the nonfluctuating model at the top of the trajectory and inside it at the bottom left.  In Fig. \ref{fig:s11_chisq_fitting_comparison}(d), where $K>0$ on the other hand, we find that our measured trajectory lies outside our best fit to the nonfluctuating model at the bottom of the trajectory and inside it at the top left.  It is important to note that this asymmetry cannot be attributed to impedance mismatches, since additional rotation of the fit to Eq. \eqref{eq:s11_linear} about the off-resonant point $S_{11} = 1$ leads to a poor fit near this point.  

\begin{figure}[!t]
  \includegraphics{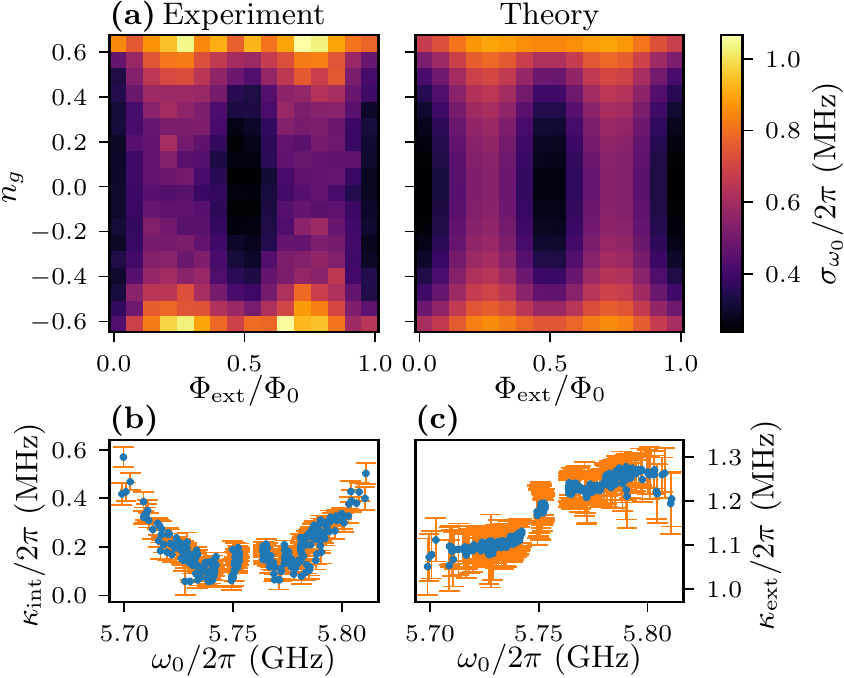}
  \caption{(a) Measured $\sigma_{\omega_{0}}(n_{g},\Phi_{\ext})$ and the best fit to Eq. \eqref{eq:renormalized_variance}, which includes the effects of frequency fluctuations due to charge noise, flux noise, and quantum noise.  In (b) and (c), $\kappa_{\mrm{int}}(n_{g},\Phi_{\ext})$ and $\kappa_{\mrm{ext}}(n_{g},\Phi_{\ext})$ are plotted parametrically as a function of $\omega_{0}(n_{g},\Phi_{\ext})$, respectively, and their confidence intervals are shown in orange.}
  \label{fig:combined_characterization}
\end{figure}

Across most of the parameter space spanned by $n_{g}$ and $\Phi_{\mathrm{ext}}$ we find that frequency fluctuations are dominated by Gaussian fluctuations in the gate and flux, rather than by quantum fluctuations.  We therefore expect Eq. \eqref{eq:avg_s11_gaussian} to be an accurate model for our measured reflection coefficients $S_{11}(\Delta)$ for most bias points, where $\sigma_{\omega_{0}}$ is renormalized by the Kerr nonlinearity according to Eq. \eqref{eq:renormalized_variance}.  This model breaks down in small regions near $(n_{g},\Phi_{\mathrm{ext}}) = (2m, k\Phi_{0}/2)$ for integers $m$ and $k$, but numerically we find that if we try fitting the Gaussian model to data generated by Eq. \eqref{eq:avg_s11_kerr} using the scale of the Kerr nonlinearity near these points ($|K|/\kappa_{\mathrm{tot}}\approx 0.3$), we extract $\sigma_{\omega_{0}}\approx K/2$ to within $20\%$ accuracy and damping rates that are accurate to within their confidence intervals.  We can therefore use the Gaussian model as a fitting function for experimental data across our entire parameter space without significantly sacrificing accuracy in our model for $\sigma_{\omega_{0}}$ or in our extracted damping rates.  Thus, to fully characterize the cCPT we measure $S_{11}(\Delta)$ at each point in parameter space and fit each measured trajectory to Eq. \eqref{eq:avg_s11_gaussian}.  This yields the best fit parameters $\sigma_{\omega_{0}}(n_{g},\Phi_{\mathrm{ext}})$, $\kappa_{\mathrm{int}}(n_{g},\Phi_{\mathrm{ext}})$, and $\kappa_{\mathrm{ext}}(n_{g},\Phi_{\mathrm{ext}})$, which we present in Fig.~\ref{fig:combined_characterization}.

As shown in Fig. \ref{fig:combined_characterization}(a), we find excellent agreement between the measured $\sigma_{\omega_{0}}(n_{g},\Phi_{\mathrm{ext}})$ and the model given by Eq. \eqref{eq:renormalized_variance}.  This agreement between theory and experiment further corroborates our model for the effect of frequency fluctuations on the reflection coefficient, and further demonstrates the significance of quantum fluctuations to the overall frequency fluctuations since they are crucial to the gate and flux dependence of $\sigma_{\omega_{0}}$.  From the best fit to this model we find the standard deviations of gate and flux fluctuations to be
\begin{equation}\label{eq:charge_and_flux_sigmas}
\begin{split}
\sigma_{n_{g}} &= (6.1 \pm 0.2) \times 10^{-3} \:\: \mrm{electrons} \\
\sigma_{\Phi_{\ext}} &= (2.80 \pm 0.02) \times 10^{-3} \:\: \Phi_{0}
\end{split}
\end{equation}
which, in general, depend on the time-scale of the individual reflection measurements used to find each value of $\sigma_{\omega_{0}}$, as we will discuss in Section \ref{sec:power_spectra_of_frequency_fluctuations}.  With these we can calculate the strength of the second order couplings between the tuning and frequency fluctuations \cite{Brock2020}
\begin{equation}
\begin{split}
|D_{n_{g}}| &= \frac{\sigma_{n_{g}}^{2}}{2}\left|\frac{\partial^{2}\omega_{0}}{\partial n_{g}^{2}}\right| < 2\pi\times 14 \: \mathrm{kHz} \\
|D_{\Phi_{\mathrm{ext}}}| &= \frac{\sigma_{\Phi_{\mathrm{ext}}}^{2}}{2}\left|\frac{\partial^{2}\omega_{0}}{\partial \Phi_{\mathrm{ext}}^{2}}\right| < 2\pi\times 10 \: \mathrm{kHz}
\end{split}
\end{equation}
where the bounds are over the full tuning range of the cCPT.  Since these are both much smaller than $\kappa_{\mathrm{tot}}$, we are justified in truncating Eq. \eqref{eq:tuning_fluctuations_frequency_coupling} at linear order.

The internal damping rate $\kappa_{\mrm{int}}(n_{g},\Phi_{\ext})$ of the cCPT, obtained by fitting our measured trajectories $S_{11}(\Delta)$ to Eq. \eqref{eq:avg_s11_gaussian}, is plotted parametrically versus $\omega_{0}(n_{g},\Phi_{\ext})$ in Fig.~\ref{fig:combined_characterization}(b).  Although $\kappa_{\mathrm{int}}$ varies somewhat at each value of $\omega_{0}$, most of its variation can be attributed to an implicit dependence on the operating frequency $\omega_{0}$ rather than an explicit dependence on $n_{g}$ and $\Phi_{\mathrm{ext}}$.  In Appendix \ref{sec:classical_circuit_model} we show that the internal damping rate takes the form
\begin{equation}
\kappa_{\mrm{int}} = \frac{4\omega_{\lambda/4}}{\pi}\alpha\ell
\end{equation}
where $\alpha$ is the attenuation constant of the cavity and $\ell$ is its length.  Thus, one possible explanation for the implicit dependence of $\kappa_{\mathrm{int}}$ on $\omega_{0}$ is the attenuation constant varying with frequency.  Another possible explanation is that the metallization between the central conductor and the ground plane (which forms the CPT) affects $\kappa_{\mathrm{int}}$ in such a way that it depends on the operating frequency.  It has previously been observed that similar metallization at high impedance points (e.g., the voltage antinode) yields an order unity change in a cavity's internal damping rate \cite{Chen2011, thesis_chen}.  As the resonant frequency is tuned, so too is its effective length and the impedance of the point at which the CPT is embedded.  It is therefore plausible that additional loss would arise as the resonant frequency is tuned further away from its bare value, which is precisely what we observe.

The external damping rate $\kappa_{\mrm{ext}}(n_{g},\Phi_{\ext})$ of the cCPT, obtained by fitting our measured trajectories $S_{11}(\Delta)$ to Eq. \eqref{eq:avg_s11_gaussian}, is plotted parametrically versus $\omega_{0}(n_{g},\Phi_{\ext})$ in Fig.~\ref{fig:combined_characterization}(c).  Clearly, the variation  in $\kappa_{\mathrm{ext}}$ can be fully attributed to an implicit dependence on the operating frequency $\omega_{0}$.  In Appendix \ref{sec:classical_circuit_model} we show that the external damping rate takes the form
\begin{equation}
\kappa_{\mrm{ext}} = \frac{4\omega_{\lambda/4}}{\pi}(\omega_{0}Z_{0}C_{c})^{2}
\end{equation}
where $Z_{0} = 50\Omega$ is the characteristic impedance of the transmission lines and $C_{c}$ is the coupling capacitance between the cavity and the external transmission line.  Although $\kappa_{\mathrm{ext}}$ depends explicitly on $\omega_{0}^{2}$ this cannot account for its measured variation, since $\kappa_{\mrm{ext}}$ deviates from its mean value by about $10\%$ while $\omega_{0}$ only varies from its bare value by about $1\%$.  Rather, we attribute the variation in $\kappa_{\mathrm{ext}}$ to the characteristic impedance $Z_{0}$ of either the cavity or its environment changing with the operating frequency.  Using our extracted value of $\kappa_{\ext}/2\pi \approx 1.2$ MHz at $\omega_{0} \approx \omega_{\lambda/4} = 2\pi\times 5.757$ GHz and the nominal value of $Z_{0} = 50\Omega$, we can solve for the coupling capacitance $C_{c} = 7.1$ fF, which is consistent with both a first principles calculation based on the geometry of the interdigitated capacitor \cite{text_gupta} and a simulation using Sonnet.

\section{Power Spectra of Frequency Fluctuations}
\label{sec:power_spectra_of_frequency_fluctuations}

To corroborate the presence and strength of frequency fluctuations, as well as shed light on their underlying sources, we next perform a direct measurement of their power spectral density (PSD).  We do so by driving the cCPT with a carrier signal on resonance and measuring the output PSD near $\omega_{0}$ using a spectrum analyzer.  This carrier signal will be modulated by the frequency fluctuations, which we assume to have PSD $S_{\Omega\Omega}(\omega)$, such that the power spectral density $S_{\mathrm{out}}$ of the output power at the plane of the sample is given by
\begin{equation}\label{eq:frequency_modulation_PSD_maintxt}
S_{\mrm{out}}(\omega_{0}\pm\omega) = \frac{2\kappa_{\mrm{ext}}^{2}}{\kappa_{\mrm{tot}}^{2}(\omega^{2} + \kappa_{\mrm{tot}}^{2}/4)}P_{\mrm{in}}S_{\Omega\Omega}(\omega)
\end{equation}
as shown in Appendix \ref{sec:frequency_modulated_cavity_response}.  This can be related to the power spectral density $S_{\mathrm{out}}^{\mathrm{SA}}$ measured by the spectrum analyzer according to
\begin{equation}
S_{\mrm{out}}^{\mathrm{SA}}(\omega_{0}\pm\omega) = \frac{2\kappa_{\mrm{ext}}^{2}}{\kappa_{\mrm{tot}}^{2}(\omega^{2} + \kappa_{\mrm{tot}}^{2}/4)}\frac{G(\omega_{0}\pm\omega)}{\eta_{\mathrm{in}}}P_{\mrm{car}}S_{\Omega\Omega}(\omega)
\end{equation}
where $G$ is the gain of the amplifier chain, $\eta_{\mathrm{in}}$ is the input attenuation, and $P_{\mathrm{car}}$ is the power of the carrier signal at the fridge input.  

As discussed in Section \ref{sec:reflection_measurements}, we can measure the ratio $G/\eta_{\mathrm{in}}$ at any frequency using the off-resonant transmission magnitude $|S_{21}|$ from the input to the output port of the fridge.  Having now determined the damping rates as well, we can measure $S_{\mrm{out}}^{\mathrm{SA}}(\omega_{0}\pm\omega)$ and invert this relationship to extract $S_{\Omega\Omega}(\omega)$.  Furthermore, $S_{\Omega\Omega}$ can be expressed in terms of the PSDs of its underlying sources as
\begin{equation}\label{eq:PSD_frequency_fluctuations}
S_{\Omega\Omega}(\omega) = \abs{\frac{\partial\omega_{0}}{\partial n_{g}}}^{2} S_{qq}(\omega) + \abs{\frac{\partial\omega_{0}}{\partial \Phi_{\ext}}}^{2} S_{\Phi\Phi}(\omega) + K^{2}S_{nn}(\omega)
\end{equation}
where $S_{qq}$ is the PSD of fluctuations in the gate charge $n_{g}$, $S_{\Phi\Phi}$ is the PSD of fluctuations in the external flux $\Phi_{\mathrm{ext}}$, and $S_{nn}$ is the PSD of quantum fluctuations in the cavity occupation $n = (X_{1}^{2} + X_{2}^{2}-2)/4$.  By carefully choosing the gate and flux biases at which we measure $S_{\Omega\Omega}$ we can isolate each of these contributions, which we can then compare with the results of Section \ref{sec:damping_rates_and_deformed_resonance_circles}.

For each of these measurements, we drive the cCPT using a carrier signal at $\omega_{0}$ with power $-60$ dBm, slightly below the single-photon level such that the cavity response is linear to good approximation.  We then measure the output power using a spectrum analyzer whose measurement window is centered at $\omega_{0}$ with a span of $100$ kHz and resolution bandwidth of $1$ Hz.  To measure the corresponding noise floor, we perform an identical measurement with the carrier signal turned off.  As expected, all measured output spectra are symmetric about $\omega_{0}$; we therefore calculate $S_{\Omega\Omega}(\omega)$ from the average of $S_{\mathrm{out}}(\omega_{0} + \omega)$ and $S_{\mathrm{out}}(\omega_{0} - \omega)$ to better resolve the fluctuations of interest from the noise floor.  For convenience we express all measured PSDs in units of frequency rather than angular frequency.  

\begin{figure}[!t]
  \includegraphics{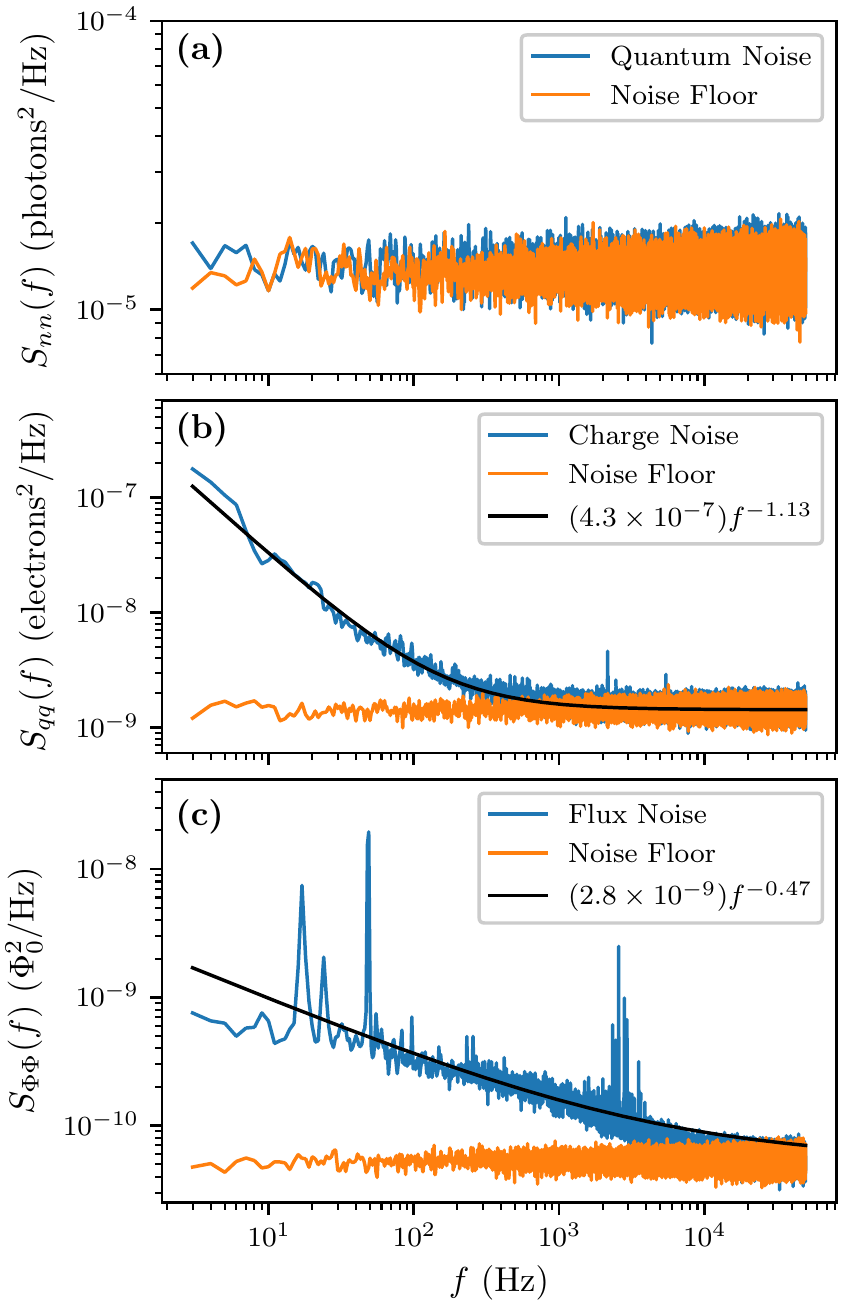}
  \caption{Measured power spectral densities of (a) quantum noise, (b) charge noise, and (c) flux noise.  The power law fits in (b) and (c) have been offset by the mean noise floors to compare with our measured data.}
  \label{fig:noise_spectra}
\end{figure}

To measure the PSD of quantum noise, $S_{nn}$, we bias the cCPT to $(n_{g},\Phi_{\ext}) = (0,0)$ where both $\partial\omega_{0}/\partial n_{g}=0$ and $\partial\omega_{0}/\partial\Phi_{\mathrm{ext}}=0$.  Thus, all power in excess of the noise floor near $\omega_{0}$ is attributable to quantum fluctuations.  Our measurement of $S_{nn}(f)$ is presented in Fig. \ref{fig:noise_spectra}(a), from which we see that we cannot resolve quantum fluctuations from the noise floor at these frequencies.  Thus, for all subsequent measurements we assume quantum fluctuations contribute negligibly to Eq. \eqref{eq:PSD_frequency_fluctuations}.

To measure the PSD of charge noise, $S_{qq}$, we bias the cCPT to $(n_{g},\Phi_{\ext}) = (0.5,0)$ where $\partial\omega_{0}/\partial\Phi_{\mathrm{ext}}=0$.  Thus, all power in excess of the noise floor near $\omega_{0}$ is attributable to fluctuations in the gate charge. Our measurement of $S_{qq}(f)$ is presented in Fig. \ref{fig:noise_spectra}(b), from which we see that it has an $f^{-\alpha}$ power law dependence.  This type of charge noise is common in solid state systems, and is believed to arise due to fluctuating two-level systems in the vicinity of the CPT island \cite{Paladino2014,Astafiev2006,Kafanov2008}.  Over the frequency range $f\lesssim 200$ Hz where the charge noise can be clearly resolved from the noise floor, we fit a power law to the charge noise in excess of the noise floor and find the best fit
\begin{equation}\label{eq:charge_noise_power_law}
S_{qq}(f) \approx (4.3\times 10^{-7})f^{-1.13} \: \mathrm{electrons}^{2}/\mathrm{Hz}.
\end{equation}
We note that our measured value of $S_{qq}(1\mathrm{Hz})$ is comparable to typical values reported in the literature for single electron transistors (SETs) \cite{Verbrugh1995,Zorin1996,Li2007}.  The total variance of fluctuations in the gate for a given measurement is obtained by integrating $S_{qq}$ over the measurement bandwidth, with lower cutoff frequency set by the inverse of the measurement time $1/\tau_{m}$ and upper cutoff frequency set by the total damping rate $\kappa_{\mathrm{tot}}/2\pi$.  To compare our PSD with the results of Section \ref{sec:damping_rates_and_deformed_resonance_circles} we use $\tau_{m} = 0.03$~s.  This corresponds to the total time spent measuring $S_{11}$, at each value of the detuning $\Delta$, for the measurements used to determine $\sigma_{n_{g}}$ and $\sigma_{\Phi_{\mathrm{ext}}}$ in that section.  Assuming the charge noise follows the power law given by Eq. \eqref{eq:charge_noise_power_law} over the full measurement bandwidth, we estimate
\begin{equation}\label{eq:sigma_ng_from_PSD}
\sigma_{n_{g}} = \sqrt{\int\limits_{1/\tau_{m}}^{\kappa_{\mrm{tot}}/2\pi}S_{qq}(f)df} = 1.3 \times 10^{-3} \:\: \mrm{electrons}
\end{equation}
which is in order-of-magnitude agreement with Eq. \eqref{eq:charge_and_flux_sigmas}.

To measure the PSD of flux noise, $S_{\Phi\Phi}$, we bias the cCPT to $(n_{g},\Phi_{\ext}) = (0,\Phi_{0}/4)$ where $\partial\omega_{0}/\partial n_{g}=0$.  Thus, all power in excess of the noise floor near $\omega_{0}$ is attributable to fluctuations in the flux threading the SQUID loop.  Our measurement of $S_{\Phi\Phi}(f)$ is presented in Fig. \ref{fig:noise_spectra}(c), from which we see that it too has an $f^{-\alpha}$ power law dependence.  This type of flux noise is ubiquitous in SQUIDs and is believed to arise from unpaired surface spins \cite{Paladino2014, Wellstood1987, Sendelbach2008, Kumar2016}.  Over the frequency range $f\lesssim 2$ kHz where the flux noise can be clearly resolved from the noise floor, we fit a power law to the flux noise in excess of the noise floor and find the best fit
\begin{equation}\label{eq:flux_noise_power_law}
S_{\Phi\Phi}(f) \approx (2.8\times 10^{-9})f^{-0.47} \: \Phi_{0}^{2}/\mathrm{Hz}.
\end{equation}
We note that our measured value of $S_{\Phi\Phi}(1\mathrm{Hz})$ is significantly larger than typical values on the order of $(\mu\Phi_{0})^{2}$/Hz found in the literature \cite{Wellstood1987, Sendelbach2008, Kumar2016}, which we attribute to the large size of our SQUID loop.  In addition, although the exponent $\alpha \approx 0.5$ in the $f^{-\alpha}$ dependence of $S_{\Phi\Phi}$ is on the low side of what has been reported in the literature, it is not unprecedented \cite{Kumar2016}.  Following the same line of reasoning as for the charge noise, we estimate the total standard deviation of flux fluctuations (over the bandwidth of the measurement used in Section \ref{sec:damping_rates_and_deformed_resonance_circles}) to be
\begin{equation}\label{eq:sigma_phi_from_PSD}
\sigma_{\Phi_{\mathrm{ext}}} = \sqrt{\int\limits_{1/\tau_{m}}^{\kappa_{\mrm{tot}}/2\pi}S_{\Phi\Phi}(f)df} = 3.1 \times 10^{-3} \:\: \Phi_{0}
\end{equation}
in good agreement with Eq. \eqref{eq:charge_and_flux_sigmas}.  The peaks in $S_{\Phi\Phi}$ from $10-100$ Hz and near $2.4$ kHz are due to a combination of electrical and vibrational interference, primarily from the pumps and compressors necessary to run our cryostat.  We estimate that this interference contributes less than $0.5\%$ to the total standard deviation of $\Phi_{\mathrm{ext}}$ over the measurement bandwidth considered.  

It is important to emphasize that due to limitations of this measurement and the analysis thereof, the values of $\sigma_{n_{g}}$ and $\sigma_{\Phi_{\mathrm{ext}}}$ obtained here should only be considered order-of-magnitude estimates for comparison with Eq. \eqref{eq:charge_and_flux_sigmas}.  First and foremost, the power spectra of interest disappear into the noise floor at frequencies several orders of magnitude smaller than $\kappa_{\mathrm{tot}}$.  Thus, to integrate $S_{qq}$ and $S_{\Phi\Phi}$ over the bandwidth of the measurements used in Section \ref{sec:damping_rates_and_deformed_resonance_circles} we have been forced to infer the high-frequency behavior of these power spectral densities from their low-frequency behavior.  We could improve on this limitation by using a near quantum-limited first stage amplifier \cite{Caves1982,Castellanos-Beltran2007,Macklin2015,White2015}, which would reduce our noise floor by an order of magnitude or more.  Second, in deriving Eq. \eqref{eq:frequency_modulation_PSD_maintxt} we have assumed that the carrier signal is on resonance at $\omega_{0}$ (see Appendix \ref{sec:frequency_modulated_cavity_response}), where the sideband power $S_{\mathrm{out}}(\omega_{0}+\omega)$ due to frequency fluctuations $S_{\Omega\Omega}(\omega)$ is maximal.  Over the course of measuring the output power at $\omega_{0}+\omega$, however, the resonant frequency will fluctuate around its average value thereby reducing both the average output power at $\omega_{0}+\omega$ and our estimate of $S_{\Omega\Omega}(\omega)$.  Since the scale of fluctuations in the resonant frequency around its average value is comparable to but not greater than $\kappa_{\mathrm{tot}}$, this will be an order unity effect.

\section{Kerr Shift}\label{sec:kerr_shift}

Since many of our measurements rely on our knowledge of the number of photons in the cavity, we next study the power-dependent shift in resonant frequency due to the Kerr nonlinearity~\cite{Yurke2006, Krantz2013, thesis_krantz}.  This will enable us to refer our input and output powers to the plane of the sample, and thereby determine the number of intracavity photons in-situ.  In Appendix \ref{sec:nonlinear_cavity_response} we show that the resonant frequency $\omega_{*}$, taken to be the frequency at which $|S_{11}|$ is minimized, is shifted from $\omega_{0}$ according to
\begin{equation}\label{eq:kerr_shift}
\omega_{*} = \omega_{0} + Kn = \omega_{0} + 4K\frac{\kappa_{\mrm{ext}}}{\kappa_{\mrm{tot}}}\frac{P_{\mathrm{in}}}{\hbar\omega_{0}\kappa_{\mrm{tot}}}
\end{equation}
where $n$ is the number of intracavity photons on resonance and $P_{\mathrm{in}}$ is the input power at the plane of the sample.  This can be expressed in terms of the input VNA power $P_{\mathrm{VNA}}$ using the input attenuation $\eta_{\mathrm{in}} = P_{\mathrm{VNA}}/P_{\mathrm{in}}$.  Thus, if we measure the slope
\begin{equation}\label{eq:kerr_shift_slope}
\frac{\partial \omega_{*}}{\partial P_{\mrm{VNA}}} = \frac{4\kappa_{\mrm{ext}}}{\hbar\omega_{0}\kappa_{\mrm{tot}}^{2}\eta_{\mrm{in}}}K
\end{equation}
we can determine $\eta_{\mathrm{in}}$ by comparing with this theoretical model, since the resonant frequency, Kerr nonlinearity, and damping rates have already been determined.  Here we have implicitly assumed that the damping rates of the cCPT do not vary with input power, which is not true in general \cite{Martinis2005,Sage2011} but is accurate for the range of input powers we use in this measurement.

\begin{figure}[!t]
  \includegraphics{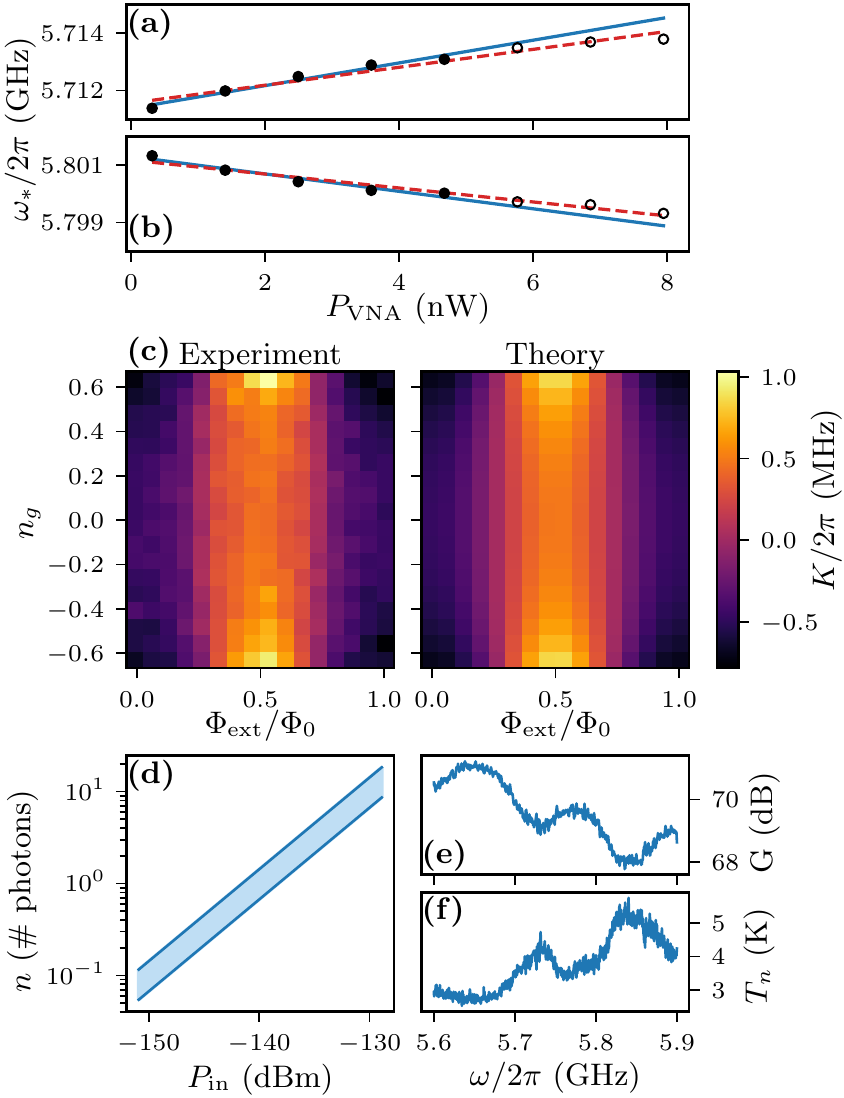}
  \caption{(a) Measurement of the Kerr-shifted resonant frequency $\omega_{*}$ (marked by black circles) as a function of $P_{\mrm{VNA}}$ at $(n_{g},\Phi_{\mrm{ext}}) = (0.56,0.53\Phi_{0})$.  The solid blue line is the best fit to the filled black circles (slope = $400$ kHz/nW), whereas the dashed red line is the best fit to both the filled and hollow black circles (slope = $310$ kHz/nW).  The uncertainty in each measured $\omega_{*}/2\pi$ is about $\pm 100$ kHz, smaller than the marker size.  (b) Same as (a), but at $(n_{g},\Phi_{\mrm{ext}}) = (0.56,0.0)$.  The slope of the solid blue line is $-300$ kHz/nW, whereas the slope of the dashed red line is $-240$ kHz/nW. (c) Measured slopes $\partial\omega_{*}/\partial P_{\mathrm{VNA}}$, scaled by a factor of $\hbar\omega_{0}\kappa_{\mathrm{tot}}^{2}\eta_{\mathrm{in}}/4\kappa_{\mathrm{ext}}$, and theoretical Kerr nonlinearity given by Eq. \eqref{eq:tunable_kerr_nonlinearity}.  (d) Number of intracavity photons $n$ at resonance as a function of $P_{\mathrm{in}}$.  (e) Gain and (f) system noise temperature of our amplifier chain as a function of frequency.}
  \label{fig:combined_kerr_shift}
\end{figure}

We perform this measurement by increasing $P_{\mathrm{VNA}}$, finding $\omega_{*}$ for each input power, and determining the slope $\partial\omega_{*}/\partial P_{\mathrm{VNA}}$.  This is done for a full period of both $n_{g}$ and $\Phi_{\mathrm{ext}}$, such that we can fit our results to Eq. \eqref{eq:kerr_shift_slope}.  The results of this measurement at two different bias points, one with positive $K$ and one with negative $K$, are shown in Fig. \ref{fig:combined_kerr_shift}(a) and \ref{fig:combined_kerr_shift}(b).  Further, we repeat this process for different ranges of input VNA powers, always starting from $-65$ dBm and incremented on a linear scale, but ending between $-56$ dBm and $-51$ dBm.  At $-56$ dBm, the cCPT is below the threshold of bistability \eqref{eq:kerr_bistability_threshold} across its full tuning range but the scale of the Kerr shift is comparable to both the cavity linewidth and the jumps in frequency between measurements at different input powers (due to slow frequency fluctuations), leading to noisiness and greater uncertainty in our measured slopes.  At $-51$ dBm, the cCPT is above the threshold of bistability across most of its tuning range, but the linear trend due to the Kerr shift is more easily resolved.  In all cases we do not observe any clear signatures of bistability, such as hysteresis and sudden jumps in $S_{11}$ as the drive frequency is swept across resonance \cite{text_nayfeh_mook}, since the total scale of frequency fluctuations over the measurement time is larger than the range of frequencies over which the response is bistable for these powers.  Empirically, we find that $\omega_{*}$ follows a linear trend with respect to $P_{\mathrm{VNA}}$ even into the bistable regime at the powers considered, but using larger maximum powers tends to yield slopes slightly smaller in magnitude leading to slightly larger input attenuations.  This effect is illustrated by the measurements shown in Figures \ref{fig:combined_kerr_shift}(a) and \ref{fig:combined_kerr_shift}(b); although the magnitude of the slopes obtained using greater maximum power are not always this much less than those obtained using lower maximum power, the trend persists on average.  This may be due to a slight increase in the internal damping rate at increasing powers, but this is difficult to determine in our case due to the complexity of simulating the nonlinear reflection coefficient, given by Eq. \eqref{eq:input_output_nonlinear_s11}, in the presence of frequency fluctuations.  Finally, it is worth noting that all of these effects, and our extracted input attenuations, are consistent across multiple cooldowns.  

We find the best agreement between theory and experiment using a maximum input VNA power of about $-54$ dBm, at which the cCPT is below the bistability threshold for all but a small region around $\Phi_{\mathrm{ext}} = \Phi_{0}/2$.  By fitting our measured slopes $\partial\omega_{*}/\partial P_{\mathrm{VNA}}$ to Eq. \eqref{eq:kerr_shift_slope} we find the input attenuation
\begin{equation}
\eta_{\mrm{in}} = (1.55 \pm 0.2)\times 10^{8} = 81.9 \pm 0.6 \: \mrm{dB}
\end{equation}
whose confidence interval is limited by the range of input attenuations extracted using different maximum input VNA powers.  In Figure \ref{fig:combined_kerr_shift}(c) we present a representative measurement of these slopes, scaled by a factor of $\hbar\omega_{0}\kappa_{\mathrm{tot}}^{2}\eta_{\mathrm{in}}/4\kappa_{\mathrm{ext}}$ so they can be compared to the Kerr nonlinearity given by Eq. \eqref{eq:tunable_kerr_nonlinearity}, and find excellent agreement between theory and experiment.  The accuracy of this connection between the Kerr nonlinearity and the measured slopes depends on the accuracy of our extracted damping rates; if we had not accounted for the presence of frequency fluctuations, our extracted input attenuation would have been skewed.  Our input attenuation is somewhat larger than its value of about $79$ dB at room temperature, contrary to our expectations since the attenuation due to our stainless steel coaxial cables should decrease slightly at cryogenic temperatures.  We believe the primary reason for this discrepancy is impedance mismatching arising at cryogenic temperatures, since all of our cables and attenuators are rated for room temperature.  Based on our room temperature measurement, we estimate that $\eta_{\mathrm{in}}$ should vary from its mean value by less than $\pm 0.2$ dB over the tuning range of $\omega_{0}$, well within its confidence interval.

Although we find excellent agreement between our measured slopes and Eq. \eqref{eq:kerr_shift_slope}, it is worth discussing two implicit assumptions of this model.  First, we have ignored the shift in $\omega_{*}$ due to frequency fluctuations \cite{Brock2020}, which would tend to increase the magnitude of our measured slopes and lead us to extract a smaller input attenuation.  Second, we have ignored the fact that frequency fluctuations reduce the average cavity occupation in the steady state, which would tend to decrease the magnitude of our measured slopes and lead us to extract a larger input attenuation.  Both of these are order unity effects that tend to cancel one another out, and modeling them rigorously would be prohibitively complex.  Thus, we have neglected them.

With this input attenuation, we can now refer our input VNA power $P_{\mathrm{VNA}}$ to the input power at the plane of the sample $P_{\mathrm{in}} = P_{\mathrm{VNA}}/\eta_{\mathrm{in}}$.  Thus, we can find the average number of photons in the cavity at resonance according to
\begin{equation}\label{eq:average_cavity_occupation}
n = \frac{4\kappa_{\ext}P_{\mrm{in}}}{\hbar\omega_{0}\kappa_{\mrm{tot}}^{2}}.
\end{equation}
which is valid in both the linear and nonlinear response regimes, as shown in Appendices \ref{sec:linear_cavity_response} and \ref{sec:nonlinear_cavity_response}.  This serves as an upper bound for the actual cavity occupation in the steady state, which will be reduced by both frequency fluctuations and non-zero detuning.  In Fig. \ref{fig:combined_kerr_shift}(d) we plot this average cavity occupation at resonance as a function of input power; the range of values at each input power is due to the variation in the damping rates as a function of operating frequency $\omega_{0}$.  

With this input attenuation we can also find the gain of our amplifier chain and the system noise referred to the plane of the sample.  We find the gain $G(\omega)$ by measuring the magnitude of the off-resonant transmission coefficient between the input and output ports of the fridge, which takes the form
\begin{equation}
\left|S_{21}^{\mathrm{VNA}}(\omega)\right| = \frac{G(\omega)}{\eta_{\mrm{in}}}
\end{equation}
as discussed in Section \ref{sec:reflection_measurements}.  Our measured gain profile is presented in Fig. \ref{fig:combined_kerr_shift}(e).  We find the system noise power spectral density $S_{\mrm{noise}}(\omega)$ by measuring the output power spectral density at room temperature $S_{\mrm{out}}(\omega)$, with no input drive, using a spectrum analyzer.  These two quantities are related to one another according to
\begin{equation}
S_{\mrm{out}}(\omega)  = G(\omega)S_{\mrm{noise}}(\omega).
\end{equation}
The power spectral density of the noise (in units of W/Hz) can be converted into a system noise temperature by dividing by the Boltzmann constant $k_{B}$.  Our measured system noise, shown in Figure \ref{fig:combined_kerr_shift}(f), is primarily due to the added noise of our first-stage cryogenic HEMT amplifier.  These results are consistent with both the specifications of the HEMT and similar results in the literature \cite{Macklin2015, White2015, Krantz2016, thesis_krantz}, thus providing additional corroboration of our extracted value for the input attenuation.

A major limitation of this method for determining the input attenuation is that we were unable to independently measure the strength of the Kerr nonlinearity, forcing us to infer its value from Eq. \eqref{eq:tunable_kerr_nonlinearity} using our extracted $E_{J}$ and $E_{C}$ given by Eq. \eqref{eq:Ej_and_Ec} and $\phi_{\mrm{zp}}$ given by Eq. \eqref{eq:phizp_val}.  This same limitation exists in other work that has used the Kerr shift (or, equivalently, the Duffing shift) to determine the power at the plane of the sample in-situ \cite{Krantz2016, Krantz2013}.  We have strong corroboration for the validity of this theoretical evaluation of $K$ from the fact that the measured $\omega_{0}(n_{g},\Phi_{\mrm{ext}})$ is in excellent agreement with theory [Fig. \ref{fig:tunable_f0_theory_exp}(a)], the slopes $\partial\omega_{*}/\partial P_{\mrm{VNA}}$ follow the same trend as our theoretical Kerr nonlinearity [Fig. \ref{fig:combined_kerr_shift}(c)], and our estimate for difference between the superconducting gaps of the island and leads of the CPT is consistent with other measurements reported in the literature.  However, a direct measurement of $K(n_{g},\Phi_{\mrm{ext}})$ would be preferable.  We might have been able to perform such a measurement if the strength of the Kerr nonlinearity exceeded the cavity linewidth \cite{Kirchmair2013}, in which case we could observe spectral signatures of $K$.  Unfortunately, in our case the observable consequences of $K$ are only sensitive to the product $P_{\mrm{in}}K$.  This is true for both the Kerr shift given by Eq. \eqref{eq:kerr_shift} and the bistability threshold given by Eq. \eqref{eq:kerr_bistability_threshold}.  Within the internal logic of this methodology our uncertainty in $K$ is determined by our confidence intervals for the best fit parameters $E_{J}$ and $E_{C}$, as well as our uncertainties $\sigma_{n_{g}}$ and $\sigma_{\Phi_{\mrm{ext}}}$ in the gate and flux bias points due to $1/f$ noise.  The total uncertainty in $K$ varies with gate and flux, but it is typically less than about $\pm 15$ kHz.

In addition, it is worth noting that a completely different method for determining the input attenuation would be possible if we were able to access the first excited state of the CPT and thereby operate it as a qubit.  Robust methods exist for calibrating the number of photons in a cavity that is coupled to a qubit via the Jaynes-Cummings interaction.  These methods make use of the Stark shift \cite{Schuster2005}, a photon-number dependent shift in the qubit frequency, the coupling strength of which can be determined independently by measuring either the phase shift of a scattered signal in the case of weak coupling \cite{thesis_palacios_laloy} or the vacuum Rabi splitting in the case of strong coupling \cite{Wallraff2004}.  Although the coupling between the qubit and cavity would be somewhat different in our case, we believe these methods could be adapted to the cCPT if not for two practical limitations.  First, due to our restricted gate range $-0.65<n_{g}<0.65$, the minimum qubit frequency we can attain with the present device is $E_{0\rightarrow 1}/h \approx 75$ GHz, which is beyond the range of frequencies we can access experimentally.  Second, we would expect our qubit (essentially a split Cooper pair box \cite{Nakamura1999}) to have a poor coherence time, making qubit spectroscopy challenging.  This could be improved if a cCPT were fabricated in the transmon regime with $E_{C}/E_{J} \ll 1$ \cite{Koch2007}, but this would then be a very different device than the one studied in the present work.

\section{Discussion}

In this work we introduce the cavity-embedded Cooper pair transistor, discuss how the CPT induces nonlinearity and tunability in the cavity, and detail the techniques used to characterize this device experimentally.  As we show, the characterization process is made significantly more complex by the presence of frequency fluctuations comparable in scale to the cavity linewidth.  Only by accounting for the effect of these fluctuations on the trajectories traced out by $S_{11}$ in the complex plane were we able to extract the true damping rates of the cCPT.  In addition, we observe the key predicted signatures of both Gaussian-distributed frequency fluctuations induced by charge and flux noise, as well as chi-square-distributed frequency fluctuations induced by quantum fluctuations of the cavity field via the Kerr nonlinearity \cite{Brock2020}.  In the latter case the signature is subtle, and we are currently investigating more direct ways of observing the consequences of these quantum frequency fluctuations.  We note that such nonlinearity-induced frequency fluctuations have also been studied in nanomechanical resonators \cite{Zhang2014,Maillet2017}, but it is unclear whether the methods employed in this context can be applied to superconducting microwave cavities with negligible thermal occupation.  

In addition to our measurements of the reflection coefficient, virtually all of our steady-state measurements are affected by frequency fluctuations, and often in ways that are difficult to model rigorously.  It may be possible to mitigate these effects by using a Pound-locking loop to stabilize the resonant frequency \cite{Lindstroem2011,Graaf2014}.  This could be achieved by measuring the error function, a signal associated with deviations of the resonant frequency from its nominal value, and using it to send a feedback signal to the gate or flux line correcting those deviations that occur slower than a cutoff frequency set by the feedback circuitry.  Since a large portion of these frequency fluctuations are due to $1/f$ noise coupling into the gate and flux coordinates, even a modest cutoff frequency on the order of $1$ kHz would significantly improve the stability of the resonant frequency.  Such a loop would also enable us to directly monitor the resonant frequency as a function of time.  

Despite these frequency fluctuations, the cCPT is a rich system with many applications.  First and foremost, it can be used for ultrasensitive electrometry at the single-photon level, as discussed experimentally \cite{Brock2021_electrometry} and theoretically \cite{Kanhirathingal2020} in our companion papers.  We have presented some evidence of this by measuring the power spectral density of frequency fluctuations due to its intrinsic charge noise using less than one photon in the cavity.  In this same vein, it is one of the key building blocks of a scheme to achieve ultra-strong optomechanical coupling at the single-photon level \cite{Rimberg2014}.  Second, we can operate it as a parametric oscillator by pumping the flux near $2\omega_{0}$, as shown in Appendix \ref{sec:cCPT_hamiltonian_parametric_pumping}, and at the same time tune the Kerr nonlinearity to zero unlike comparable systems \cite{Wilson2010,Krantz2013}.  Since the Kerr nonlinearity (equivalent to the Duffing nonlinearity within the rotating wave approximation) is generally understood to be the primary factor limiting the steady-state amplitude of parametric oscillation \cite{Dykman1998,Wustmann2013}, this would mean exploring a new regime of parametric resonance and the dynamical Casimir effect \cite{Wilson2011, Laehteenmaeki2013,Svensson2018}.  Lastly, the use of parametric pumping could enable novel charge detection schemes analogous to those that have been used for single-shot readout of superconducting qubits \cite{Lin2014, Krantz2016}.

\begin{acknowledgments}
We thank R. McDermott for assistance in producing our Nb cavities.  The sample was fabricated at Dartmouth College and the Harvard Center for Nanoscale Systems.  B.L.B., S.K., B.T., W.F.B., and A.J.R. were supported by the National Science Foundation under Grant No. DMR-1807785.  J.L. was supported by the Army Research Office under Grant No. W911NF-13-1-0377.  M.P.B. was supported by the National Science Foundation under Grant No. DMR-1507383.  
\end{acknowledgments}

\appendix

\section{Classical Circuit Model}\label{sec:classical_circuit_model}
\begin{figure}[!t]
  \includegraphics{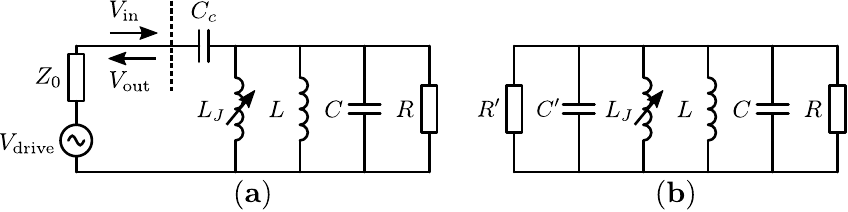}
  \caption{Schematics representing the cCPT and its external environment.  In (a), the external elements $Z_{0}$ and $C_{c}$ are in series with the resonator, as they are in the experiment, and the plane of the sample is marked by a dashed line.  In (b), the external elements are represented by their parallel equivalents: resistance $R'$ and capacitance $C'$.}
  \label{fig:rlc_circuit}
\end{figure}

Near its fundamental frequency, the input impedance of our quarter-wavelength cavity is approximately \cite{text_pozar}
\begin{equation}
Z_{\mrm{in}} \approx \frac{Z_{0}}{\alpha\ell + i\pi\Delta/2\omega_{\lambda/4}^{(0)}}
\end{equation}
where $\Delta = \omega-\omega_{\lambda/4}^{(0)}$ is the detuning from the fundamental frequency of the bare cavity, $\alpha$ is the cavity's attenuation constant, $\ell$ is its length, $Z_{0}$ is its characteristic impedance, and we have assumed both low loss ($\alpha\ell\ll 1$) and small detuning ($\Delta \ll \omega_{\lambda/4}^{(0)}$).  This input impedance is equivalent to that of a parallel RLC circuit
\begin{equation}
Z_{\mathrm{in}} \approx \frac{1}{(1/R) + 2i\Delta C}
\end{equation}
near its resonant frequency, with circuit parameters
\begin{align}
R &= \frac{Z_{0}}{\alpha\ell} \\
L &= \frac{4 Z_{0}}{\pi \omega_{\lambda/4}^{(0)}} \\
C &= \frac{\pi}{4 Z_{0}\omega_{\lambda/4}^{(0)}}. \label{eq:RLC_effective_capacitance}
\end{align}
As shown in Section \ref{sec:cCPT}, the CPT behaves as a tunable inductance $L_{J}$ in parallel with this circuit.  We couple to this system with a series capacitance $C_{c}$, and we measure it by connecting it to a network analyzer with the same characteristic impedance $Z_{0} = 50\Omega$.  The loaded cCPT can therefore be approximated by the circuit shown in Figure \ref{fig:rlc_circuit}(a).

To simplify our analysis \cite{thesis_sillanpaa}, we can replace the series combination of $Z_{0}$ and $C_{c}$ by an equivalent combination of $R'$ and $C'$ in parallel, as shown in Figure \ref{fig:rlc_circuit}(b).  We find $R'$ and $C'$ with the constraint
\begin{equation}
Z_{0}-\frac{i}{\omega C_{c}} = \left(\frac{1}{R'}+i\omega C'\right)^{-1}
\end{equation}
that yields the solutions
\begin{equation}
\begin{split}
R' &= Z_{0}\left(1+\frac{1}{\omega^{2}Z_{0}^{2}C_{c}^{2}}\right) \approx \frac{1}{\omega^{2}Z_{0}C_{c}^{2}} \\
C' &= C_{c}\left(\frac{1}{1+\omega^{2}Z_{0}^{2}C_{c}^{2}}\right) \approx C_{c}
\end{split}
\end{equation}
where the approximate expressions were obtained using $C_{c}/C \ll 1$ and $\Delta \ll \omega_{\lambda/4}^{(0)}$.  We therefore see that the bare resonant frequency of the cavity is renormalized by the coupling capacitance $C_{c}$ according to
\begin{equation}
\omega_{\lambda/4} = \frac{\omega_{\lambda/4}^{(0)}}{\sqrt{1+C_{c}/C}}
\end{equation}
and the resonant frequency $\omega_{0}$ of the cCPT can be tuned via the Josephson inductance $L_{J}$ according to
\begin{equation}
\omega_{0} = \omega_{\lambda/4}\sqrt{1+\frac{L}{L_{J}}}.
\end{equation}

Drawing the circuit in this way also allows us to easily analyze the damping rates of our system, since the internal damping rate $\kappa_{\mrm{int}}$ is due to dissipation in $R$ and the external damping rate $\kappa_{\mrm{ext}}$ is due to dissipation in $R'$.  These damping rates at resonance are defined by \cite{text_pozar}
\begin{equation}
\kappa \equiv \frac{\textrm{energy loss/second}}{\textrm{average energy stored}}
\end{equation}
where the average energy stored in a cycle is a combination of the electrical energy $W_{e} =  (C+C_{c})|V_{\mrm{in}}|^{2}/4$ and the magnetic energy $W_{m}$ (which equals $W_{e}$ at resonance), and the average energy loss per second in a cycle due to the resistance $R_{\mrm{loss}}$ is $P_{\mrm{loss}} = |V_{\mrm{in}}|^{2}/2R_{\mrm{loss}}$.  Putting these expressions together, our internal and external damping rates take the form
\begin{equation}
\begin{split}
\kappa_{\mrm{int}} &= \frac{1}{R\left(C+C_{c}\right)} \approx \frac{4\omega_{\lambda/4}}{\pi}\alpha\ell\\
\kappa_{\mrm{ext}} &= \frac{\omega_{0}^{2}Z_{0}C_{c}^{2}}{\left(C+C_{c}\right)} \approx \frac{4\omega_{\lambda/4}}{\pi}(\omega_{0}Z_{0}C_{c})^{2}
\end{split}
\end{equation}
where we have applied the same approximations as before.

We now analyze the reflection coefficient of the cCPT at the plane of the sample, indicated by the dotted line in Fig. \ref{fig:rlc_circuit}(a).  In this case we consider everything other than $Z_{0}$ to be part of the resonant circuit, including the series coupling capacitance $C_{c}$.  The impedance of the cCPT is then
\begin{equation}
Z_{\mrm{cCPT}} = \frac{1}{i\omega C_{c}} + \left(\frac{1}{R} + \frac{1}{i\omega L_{\mathrm{tot}}} + i\omega C\right)^{-1}
\end{equation}
where $L_{\mathrm{tot}}$ is the parallel combination of $L$ and $L_{J}$, which gives rise to the reflection coefficient
\begin{equation}
S_{11} = \frac{V_{\mrm{out}}}{V_{\mrm{in}}} = \frac{Z_{\mrm{cCPT}} - Z_{0}}{Z_{\mrm{cCPT}} + Z_{0}}.
\end{equation}
Plugging in and using the same approximations as earlier, we find
\begin{equation}
S_{11} = \frac{1 - x^{2} + ix\left[\cfrac{\kappa_{\mathrm{int}}}{\omega_{0}}-\cfrac{\kappa_{\mathrm{ext}}}{\omega_{0}}\left(\cfrac{C+C_{c}}{C_{c}}\right) + \cfrac{\kappa_{\mathrm{ext}}}{\omega_{0}}\cfrac{C}{C_{c}}x^{2}\right]}{1 - x^{2} + ix\left[\cfrac{\kappa_{\mathrm{int}}}{\omega_{0}} + \cfrac{\kappa_{\mathrm{ext}}}{\omega_{0}}\left(\cfrac{C+C_{c}}{C_{c}}\right) - \cfrac{\kappa_{\mathrm{ext}}}{\omega_{0}}\cfrac{C}{C_{c}}x^{2}\right]}
\end{equation}
where $x = \omega/\omega_{0}$ and $\omega_{0} = 1/\sqrt{L_{\mathrm{tot}}(C+C_{c})}$.  We next rewrite $x = 1 + \delta x$ and expand to lowest order in $\delta x = \Delta/\omega_{0} \ll 1$ to find
\begin{equation}
S_{11} = \frac{-2\delta x + i\left[\cfrac{\kappa_{\mathrm{int}}}{\omega_{0}} - \cfrac{\kappa_{\mathrm{ext}}}{\omega_{0}} + 2\cfrac{\kappa_{\mathrm{ext}}}{\omega_{0}}\cfrac{C}{C_{c}}\delta x\right]}{-2\delta x + i\left[\cfrac{\kappa_{\mathrm{int}}}{\omega_{0}} + \cfrac{\kappa_{\mathrm{ext}}}{\omega_{0}} - 2\cfrac{\kappa_{\mathrm{ext}}}{\omega_{0}}\cfrac{C}{C_{c}}\delta x\right]}
\end{equation}
where we have used the fact that $\omega_{0}\gg \kappa_{\mathrm{int}},\kappa_{\mathrm{ext}}$.  Finally, although $C_{c}/C \ll 1$, we also have $\kappa_{\mathrm{ext}}C/\omega_{0}C_{c} \ll 1$ such that 
\begin{equation}\label{eq:circuit_model_s11}
S_{11} = \frac{\Delta - i(\kappa_{\mathrm{int}} - \kappa_{\mathrm{ext}})/2}{\Delta - i(\kappa_{\mathrm{int}} + \kappa_{\mathrm{ext}})/2}
\end{equation}
where we have multiplied through by $-\omega_{0}/2$.

\section{Linear Cavity Response}\label{sec:linear_cavity_response}


To model the response of a cavity to an external drive and in the presence of damping, we use input-output theory \cite{Gardiner1985}.  In our case, the quantum Langevin equation for the cavity mode operator $a$ takes the form
\begin{equation}\label{eq:quantum_langevin_equation}
\dot{a} = \frac{i}{\hbar}\lbrack H, a \rbrack - \frac{\kappa_{\mathrm{tot}}}{2} a+ \sqrt{\kappa_{\mathrm{ext}}}a_{\mathrm{in}}(t) + \sqrt{\kappa_{\mathrm{int}}}b_{\mathrm{in}}(t)
\end{equation}
where $H$ is the cavity Hamiltonian, $\kappa_{\mathrm{tot}} = \kappa_{\mathrm{int}}+\kappa_{\mathrm{ext}}$ is the total damping rate, $a_{\mathrm{in}}$ is the input field in the transmission line coupled to the cavity via the capacitance $C_{c}$, and $b_{\mathrm{in}}$ is the noisy, zero-mean input field due to the internal loss channel.  The output field $a_{\mathrm{out}}$ in the transmission line is related to the input field and intra-cavity field according to
\begin{equation}\label{eq:input_output_relationship}
a_{\mathrm{out}} = a_{\mathrm{in}} - \sqrt{\kappa_{\mathrm{ext}}}a.
\end{equation}
A similar relationship holds between $b_{\mathrm{in}}$ and $b_{\mathrm{out}}$, but these are inaccessible experimentally.

Here we consider driving the cavity at sufficiently low powers such that its response is linear to good approximation.  We therefore take the Hamiltonian to be $H = \hbar\omega_{0}a^{\dagger}a$, and Eq.~\eqref{eq:quantum_langevin_equation} takes the form
\begin{equation}\label{eq:linear_cavity_eom}
\dot{a} = \left(-i\omega_{0} - \frac{\kappa_{\mathrm{tot}}}{2}\right)a+ \sqrt{\kappa_{\mathrm{ext}}}a_{\mathrm{in}}(t) + \sqrt{\kappa_{\mathrm{int}}}b_{\mathrm{in}}(t).
\end{equation}
Assuming a sinusoidal input drive of the form $\langle a_{\mathrm{in}}(t) \rangle = \alpha_{\mathrm{in}}e^{-i\omega t}$, the average steady-state cavity response takes the form $\langle a(t) \rangle = \alpha e^{-i\omega t}$.  Taking the ensemble average of Eq.~\eqref{eq:linear_cavity_eom} and plugging in these expressions we can solve for the intra-cavity amplitude
\begin{equation}
\alpha = \frac{\sqrt{\kappa_{\mathrm{ext}}}\alpha_{\mathrm{in}}}{-i\Delta + \kappa_{\mathrm{tot}}/2}
\end{equation}
where $\Delta = \omega - \omega_{0}$ is the detuning of the drive from resonance.  The average number of intra-cavity photons $n = |\alpha|^{2}$ in the steady state therefore takes the form
\begin{equation}\label{eq:linear_intracavity_photons}
n = \frac{\kappa_{\mathrm{ext}}P_{\mathrm{in}}/\hbar\omega}{\Delta^{2} + \kappa_{\mathrm{tot}}^{2}/4}
\end{equation}
where we have introduced the input power $P_{\mathrm{in}} = \hbar\omega |\alpha_{\mathrm{in}}|^{2}$.

Plugging these results into Eq.~\eqref{eq:input_output_relationship}, we can solve for the average steady-state amplitude of the output field
\begin{equation}
\alpha_{\mathrm{out}} = \frac{\Delta + i(\kappa_{\mathrm{int}} - \kappa_{\mathrm{ext}})/2}{\Delta + i(\kappa_{\mathrm{int}} + \kappa_{\mathrm{ext}})/2}\alpha_{\mathrm{in}}.
\end{equation}
The linear reflection coefficient can now be found from the relationship $S_{11} = (\alpha_{\mathrm{out}}/\alpha_{\mathrm{in}})^{*}$, where the complex conjugate is taken so that its phase corresponds to a counter-clockwise rotation in phase space of the output quadrature operators relative to the inputs.  In this case the reflection coefficient takes the form
\begin{equation}\label{eq:input_output_linear_s11}
S_{11} = \frac{\Delta - i(\kappa_{\mathrm{int}} - \kappa_{\mathrm{ext}})/2}{\Delta - i(\kappa_{\mathrm{int}} + \kappa_{\mathrm{ext}})/2}
\end{equation}
in agreement with Eq.~\eqref{eq:circuit_model_s11} derived from the classical circuit model.

\section{Nonlinear Cavity Response}\label{sec:nonlinear_cavity_response}

We now analyze the response of a cavity with a Kerr nonlinearity to a sinusoidal input drive.  Our Hamiltonian in this case is
\begin{equation}\label{eq:kerr_hamiltonian_rwa}
H = \hbar\omega_{0}a^{\dagger}a + \frac{1}{2}\hbar K a^{\dagger 2}a^{2}
\end{equation}
such that the quantum Langevin equation of Eq.~\eqref{eq:quantum_langevin_equation} takes the form
\begin{equation}\label{eq:nonlinear_cavity_eom}
\begin{split}
\dot{a} = \left[-i(\omega_{0} + K a^{\dagger}a) - \frac{\kappa_{\mathrm{tot}}}{2} \right]a &+ \sqrt{\kappa_{\mrm{ext}}}a_{\mrm{in}}(t) \\
&+ \sqrt{\kappa_{\mrm{int}}}b_{\mrm{in}}(t).
\end{split}
\end{equation}
As in the previous section, we assume a sinusoidal input drive of the form $\langle a_{\mathrm{in}}(t) \rangle = \alpha_{\mathrm{in}}e^{-i\omega t}$ such that the average steady-state cavity response takes the form $\langle a(t) \rangle = \alpha e^{-i\omega t}$.  Plugging back into Eq.~\eqref{eq:nonlinear_cavity_eom}, we find 
\begin{equation}\label{eq:kerr_alpha_response}
\left[-i\left(\Delta - K\left|\alpha\right|^{2}\right)+\frac{\kappa_{\mathrm{tot}}}{2}\right]\alpha = \sqrt{\kappa_{\mrm{ext}}}\alpha_{\mrm{in}}
\end{equation}
where $\Delta = \omega - \omega_{0}$ is the detuning of the drive from resonance.  To solve this nonlinear equation for the intra-cavity field amplitude $\alpha$, we multiply both sides by their complex conjugates and find
\begin{equation}\label{eq:kerr_response}
K^{2}n^{3} - 2\Delta K n^{2} + \left[\Delta^{2} + \frac{\kappa_{\mrm{tot}}^{2}}{4}\right]n - \kappa_{\mrm{ext}}\frac{P_{\mrm{in}}}{\hbar\omega_{0}} = 0
\end{equation}
where $n = \abs{\alpha}^{2}$ is the average number of intra-cavity photons, $P_{\mrm{in}} = \hbar\omega\abs{\alpha_{\mrm{in}}}^{2}$ is the power at the input port of the cavity, and we have used $1/\omega\approx1/\omega_{0}$ since $\Delta\ll\omega_{0}$.  

In general, this cubic equation for the cavity response has three solutions; if only one solution is real then it is stable and the unique physical solution, whereas if all three solutions are real then one solution will be unstable and there will be bistability in the cavity response as a function of detuning \cite{Yurke2006}.  The bifurcation between these two regimes occurs when the response curve $n(\Delta)$ becomes vertical such that $\partial\Delta/\partial n = 0$.  Differentiating Eq.~\eqref{eq:kerr_response} with respect to $n$ and evaluating at the critical number of photons $n_{c}$ where this bifurcation occurs, we find
\begin{equation}
n_{c} = \frac{2\Delta}{3K}\left(1\pm \frac{1}{2}\sqrt{1-\frac{3\kappa_{\mathrm{tot}}^{2}}{4\Delta^{2}}}\right).
\end{equation}
Since the number of photons $n_{c}$ must be real and positive this equation implies that there is also a critical detuning 
\begin{equation}
\Delta_{c} = \mathrm{sign}(K)\frac{\sqrt{3}\kappa_{\mathrm{tot}}}{2}
\end{equation}
at which the onset of bistability occurs, such that
\begin{equation}
n_{c} = \frac{\sqrt{3}}{3}\frac{\kappa_{\mathrm{tot}}}{|K|}.
\end{equation}
Plugging these back into Eq.~\eqref{eq:kerr_response}, we can solve for the critical input power
\begin{equation}\label{eq:kerr_bistability_threshold}
P_{c} = \frac{\sqrt{3}}{9}\frac{\hbar\omega_{0}\kappa_{\mathrm{tot}}^{3}}{|K|\kappa_{\mathrm{ext}}}
\end{equation}
above which the response is bistable for a range of detunings near $\Delta_{c}$.  

Below this bistability threshold, we can solve Eq.~\eqref{eq:kerr_response} for $n$ uniquely and plug back into Eq.~\eqref{eq:kerr_alpha_response} to find the phase of $\alpha$ self-consistently.  From here we can use Eq.~\eqref{eq:input_output_relationship} to solve for the average steady state output field $\alpha_{\mathrm{out}}$, and thus the reflection coefficient $S_{11} = (\alpha_{\mathrm{out}}/\alpha_{\mathrm{in}})^{*}$ as in the previous section.  Doing so, we find the nonlinear reflection coefficient takes the form
\begin{equation}\label{eq:input_output_nonlinear_s11}
S_{11} = \frac{\Delta - Kn - i(\kappa_{\mathrm{int}} - \kappa_{\mathrm{ext}})/2}{\Delta - Kn - i(\kappa_{\mathrm{int}} + \kappa_{\mathrm{ext}})/2}
\end{equation}
where $n$ depends on the detuning $\Delta$ according to Eq. \eqref{eq:kerr_response}.  This expression is equivalent to the linear case of Eq.~\eqref{eq:input_output_linear_s11} except the detuning is shifted according to $\Delta \rightarrow \Delta - Kn$.  In this case the resonant frequency, taken to be the minimum of $|S_{11}|$, occurs at the detuning $\Delta_{*} = Kn$.  Plugging this result back into Eq.~\eqref{eq:kerr_response} we can solve for the shifted resonant frequency
\begin{equation}
\omega_{*} = \omega_{0} + 4K\frac{\kappa_{\mrm{ext}}}{\kappa_{\mrm{tot}}}\frac{P_{\mathrm{in}}}{\hbar\omega_{0}\kappa_{\mrm{tot}}}
\end{equation}
in terms of the input power.  Finally, it is worth noting that the number of photons in the cavity at this shifted resonant frequency takes the form
\begin{equation}
n = \frac{4\kappa_{\mathrm{ext}}P_{\mathrm{in}}}{\hbar\omega_{0}\kappa_{\mathrm{tot}}^{2}}
\end{equation}
which is equivalent to Eq.~\eqref{eq:linear_intracavity_photons} evaluated at resonance.

\section{Frequency-modulated Cavity Response}\label{sec:frequency_modulated_cavity_response}
For this analysis we go back to the cavity equation of motion
\begin{equation}
\dot{a} = -i\omega_{0}(t)a - \frac{\kappa_{\mrm{tot}}}{2}a + \sqrt{\kappa_{\mrm{ext}}}a_{\mrm{in}}
\end{equation}
where we consider modulating the resonant frequency such that
\begin{equation}
\omega_{0}(t) = \omega_{0} + \Omega\cos(\omega_{m}t).
\end{equation}
We treat the system semiclassically by taking the expectation value of this equation of motion and assuming a sinusoidal drive $\expval{a_{\mrm{in}}(t)}=\alpha_{\mrm{in}}e^{-i\omega t}$, such that
\begin{equation}
\dot{\alpha} = \left[-i\omega_{0} - i\Omega\cos(\omega_{m}t) - \frac{\kappa_{\mrm{tot}}}{2}\right] \alpha + \sqrt{\kappa_{\mrm{ext}}}\alpha_{\mrm{in}}e^{-i\omega t}.
\end{equation}
We now make an ansatz of the form
\begin{equation}
\alpha(t) = A(t)\exp\left[-i\omega_{0}t - i\frac{\Omega}{\omega_{m}}\sin(\omega_{m}t) - \frac{\kappa_{\mrm{tot}}}{2}t \right]
\end{equation}
and find the equation of motion for the amplitude $A(t)$
\begin{equation}
\dot{A} = \sqrt{\kappa_{\mrm{ext}}}\alpha_{\mrm{in}}\exp\left[-i\Delta t + i\frac{\Omega}{\omega_{m}}\sin(\omega_{m}t) + \frac{\kappa_{\mrm{tot}}}{2}t \right].
\end{equation}
where we have introduced the detuning $\Delta = \omega-\omega_{0}$.

To solve this differential equation we use the Jacobi-Anger expansion
\begin{equation}
e^{iz\sin(x)} = \sum\limits_{n=-\infty}^{\infty}J_{n}(z)e^{inx}
\end{equation}
where $J_{n}$ is the nth Bessel function of the first kind.  Plugging this identity into our differential equation, we find
\begin{equation}
\dot{A} = \sqrt{\kappa_{\mrm{ext}}}\alpha_{\mrm{in}}\sum\limits_{n=-\infty}^{\infty}J_{n}\left(\frac{\Omega}{\omega_{m}}\right) e^{-i(\Delta-n\omega_{m})t + \kappa_{\mrm{tot}}t/2}.
\end{equation}
This equation can be integrated directly, yielding the solution
\begin{equation}
A(t) = \sqrt{\kappa_{\mrm{ext}}}\alpha_{\mrm{in}}\sum\limits_{n=-\infty}^{\infty}J_{n}\left(\frac{\Omega}{\omega_{m}}\right)\frac{e^{-i(\Delta-n\omega_{m})t + \kappa_{\mrm{tot}}t/2}}{-i(\Delta-n\omega_{m}) + \kappa_{\mrm{tot}}/2}
\end{equation}
where we have dropped the constant $A(0)$ since we are interested in the steady state, rather than the transient response.  Plugging back into our ansatz, we arrive at the solution for the intracavity expectation value
\begin{equation}
\begin{split}
\alpha(t) = &\sqrt{\kappa_{\mrm{ext}}}\alpha_{\mrm{in}} \\
&\times\sum\limits_{n=-\infty}^{\infty}J_{n}\left(\frac{\Omega}{\omega_{m}}\right)\frac{e^{-i(\omega-n\omega_{m})t - i(\Omega/\omega_{m})\sin(\omega_{m}t)}}{-i(\Delta-n\omega_{m}) + \kappa_{\mrm{tot}}/2} .
\end{split}
\end{equation}
Applying the Jacobi-Anger expansion again, we find
\begin{equation}
\begin{split}
\alpha(t) = \sqrt{\kappa_{\mrm{ext}}}\alpha_{\mrm{in}}&\sum\limits_{n=-\infty}^{\infty}J_{n}\left(\frac{\Omega}{\omega_{m}}\right)\frac{\exp\left[-i(\omega-n\omega_{m})t \right]}{-i(\Delta-n\omega_{m}) + \frac{\kappa_{\mrm{tot}}}{2}} \\
\times & \sum\limits_{k=-\infty}^{\infty}J_{k}\left(\frac{\Omega}{\omega_{m}}\right) \exp\left[-ik(\omega_{m}t) \right].
\end{split}
\end{equation}

We next simplify this expression by evaluating at $\Delta=0$ (where the response is maximized) and expanding to linear order in $\Omega/\omega_{m}$ (which we assume to be much less than one for a weak frequency modulation amplitude), yielding the expression
\begin{equation}
\begin{split}
\alpha(t) = \sqrt{\kappa_{\mrm{ext}}}\alpha_{\mrm{in}}\Biggl[\frac{2}{\kappa_{\mrm{tot}}}e^{-i\omega_{0} t} -\frac{i\Omega e^{-i(\omega_{0}+\omega_{m})t}}{\kappa_{\mrm{tot}}(-i\omega_{m} + \kappa_{\mrm{tot}}/2)} \quad & \\
-\frac{i\Omega e^{-i(\omega_{0}-\omega_{m})t}}{\kappa_{\mrm{tot}}(i\omega_{m} + \kappa_{\mrm{tot}}/2)} \Biggr].&
\end{split}
\end{equation}
Finally, we solve for the output field 
\begin{equation}
\begin{split}
\alpha_{\mrm{out}} = \alpha_{\mrm{in}}\Biggl[\left(1 - 2\frac{\kappa_{\mrm{ext}}}{\kappa_{\mrm{tot}}}\right)e^{-i\omega_{0} t} +\frac{i\kappa_{\mrm{ext}}\Omega e^{-i(\omega_{0}+\omega_{m})t}}{\kappa_{\mrm{tot}}(-i\omega_{m} + \kappa_{\mrm{tot}}/2)} \; &\\
+\frac{i\kappa_{\mrm{ext}}\Omega e^{-i(\omega_{0}-\omega_{m})t}}{\kappa_{\mrm{tot}}(i\omega_{m} + \kappa_{\mrm{tot}}/2)} \Biggr] &
\end{split}
\end{equation}
using the input-output relation $a_{\mrm{out}} = a_{\mrm{in}} - \sqrt{\kappa_{\mrm{ext}}}a$.

Since the ingoing and outgoing voltages are proportional to $a_{\mrm{in}}$ and $a_{\mrm{out}}$, the amplitude of the outgoing voltage at each of the sidebands is given by
\begin{equation}
V_{\mrm{out}}(\omega_{0}\pm\omega_{m}) = \frac{\kappa_{\mrm{ext}}\Omega}{\kappa_{\mrm{tot}}\sqrt{\omega_{m}^{2} + \kappa_{\mrm{tot}}^{2}/4}}V_{\mrm{in}}
\end{equation}
such that we can simply read off the power at the sidebands
\begin{equation}
P_{\mrm{out}}(\omega_{0}\pm\omega_{m}) = \frac{\kappa_{\mrm{ext}}^{2}\Omega^{2}}{\kappa_{\mrm{tot}}^{2}(\omega_{m}^{2} + \kappa_{\mrm{tot}}^{2}/4)}P_{\mrm{in}}.
\end{equation}
If the frequency modulation is instead a noisy signal with power spectral density $S_{\Omega\Omega}(\omega)$ rather than a pure tone, the output power spectral density $S_{\mathrm{out}}$ will take the form
\begin{equation}\label{eq:frequency_modulation_PSD}
S_{\mrm{out}}(\omega_{0}\pm\omega) = \frac{2\kappa_{\mrm{ext}}^{2}}{\kappa_{\mrm{tot}}^{2}(\omega^{2} + \kappa_{\mrm{tot}}^{2}/4)}P_{\mrm{in}}S_{\Omega\Omega}(\omega)
\end{equation}
where we have picked up an extra factor of two since the power spectral density is defined in terms of RMS modulation amplitude per unit bandwidth.  Thus, if we drive the cavity at its resonant frequency we can extract the power spectral density of frequency noise $S_{\Omega\Omega}(\omega)$ by measuring the output power spectral density $S_{\mathrm{out}}(\omega_{0}\pm\omega)$ at the corresponding sideband frequencies.  

\section{Parametrically-pumped cCPT Hamiltonian}\label{sec:cCPT_hamiltonian_parametric_pumping}
For completeness, here we derive the additional term that arises in the cCPT Hamiltonian when the flux line is parametrically pumped near $2\omega_{0}$.  We start from Eq. \eqref{eq:cCPT_expansion_full}, which takes the form
\begin{equation}
H = \frac{Q^{2}}{2C} + \frac{\Phi^{2}}{2L}
+ \sum_{k=0}^{\infty}\frac{1}{k!}\partial^{k}_{\phi}E_{CPT}(n_{g},\Phi_{\ext})\left(\frac{2\pi\Phi}{\Phi_{0}}\right)^{k}.
\end{equation}
If the flux is pumped at $\omega_{p}\sim 2\omega_{0}$ such that
\begin{equation}
\Phi_{\mrm{ext}} \rightarrow \Phi_{\mrm{ext}} + \delta\Phi\cos(\omega_{p}t)
\end{equation}
for $\delta\Phi\ll\Phi_{0}$, then $E_{CPT}$ can be expanded to linear order in $\delta\Phi$ yielding the additional term
\begin{equation}
\begin{split}
H_{\mrm{pump}} =\: &  \delta\Phi\cos(\omega_{p}t) \\
&\times \frac{\partial}{\partial\Phi_{\mrm{ext}}}\sum_{k=0}^{\infty}\frac{\phi_{\mrm{zp}}^{k}}{k!}\partial^{k}_{\phi}E_{CPT}(n_{g},\Phi_{\ext})\left(a+a^{\dagger}\right)^{k}
\end{split}
\end{equation}
in the Hamiltonian, where we've expressed the cavity flux coordinate $\Phi$ in terms of mode operators $a$ and $a^{\dagger}$ using Eq. \eqref{eq:cavity_mode_operator}.  Since the mode operators $a$ and $a^{\dagger}$ tend to oscillate as $e^{-i\omega_{0}t}$ and $e^{i\omega_{0}t}$, respectively, all the even-$k$ terms in this expansion will have components that oscillate as $e^{\pm2i\omega_{0}t}$, which will be made stationary by the flux pump near $2\omega_{0}$.  

Thus, if we make the rotating wave approximation as in Sec. \ref{sec:cCPT_Hamiltonian} by dropping all terms that oscillate rapidly in the frame rotating at $\omega_{p}/2\sim\omega_{0}$, this pump term can be approximated
\begin{equation}
H_{\mrm{pump}} \approx \frac{\hbar}{4} \frac{\partial\omega_{0}}{\partial\Phi_{\mrm{ext}}} \delta\Phi \left(e^{i\omega_{p}t}a^{2} + e^{-i\omega_{p}t}a^{\dagger 2}\right)
\end{equation}
to leading order in $\phi_{\mrm{zp}}$ (neglecting the constant $k=0$ term), where we've expressed the cosine as a complex exponential and written $\partial_{\phi}^{2}E_{CPT}$ in terms of $\omega_{0}$ using Eq. \eqref{eq:tunable_resonant_frequency}.  The total Hamiltonian can therefore be cast in the standard form of a parametric oscillator with a Kerr nonlinearity \cite{Wang2019}
\begin{equation}
H = \hbar\omega_{0} a^{\dagger}a + \frac{1}{2}\hbar Ka^{\dagger 2}a^{2} + \frac{1}{2}\hbar\epsilon\left(e^{i\omega_{p}t} a^{2} + e^{-i\omega_{p}t} a^{\dagger 2}\right)
\end{equation}
where the pump strength $\epsilon$ is given by
\begin{equation}
\epsilon = \frac{1}{2}\frac{\partial\omega_{0}}{\partial\Phi_{\mrm{ext}}} \delta\Phi,
\end{equation}
which depends on both gate and flux.  An analysis of the dynamics of the parametrically-pumped cCPT is left for future work.

\section{Self-inductance of the SQUID Loop}\label{sec:SQUID_self_inductance}

In Eq. \eqref{eq:SQUID_loop_superconducting_phase} we neglected the contribution of the SQUID loop's self-inductance $L_{\mrm{loop}}$ to the total flux through the loop.  In general, however, if a current $I_{\mrm{circ}}$ is circulating around the loop this will induce a flux $L_{\mrm{loop}}I_{\mrm{circ}}$ through it.  This contribution is negligible in our case if it is significantly less than the magnetic flux quantum, such that $L_{\mrm{loop}}|I_{\mrm{circ}}|/\Phi_{0}\ll 1$.  Although the SQUID loop does have a relatively large self-inductance due to its size, the current flowing through the CPT in its ground state is sufficiently small that this condition holds for all values of the external flux $\Phi_{\mrm{ext}}$ over the gate range $-0.65<n_{g}<0.65$.

Treating the loop as rectangular (approximately $8$ mm $\times$ $5$ $\mu$m), we conservatively estimate its self-inductance to be $L_{\mrm{loop}}\lesssim 10$ nH.  To estimate the circulating current we first note that it is simply the average current flowing through the JJs, which can be written
\begin{equation}
I_{\mrm{circ}} = \frac{1}{2}I_{c}\left\langle \sin(\varphi_{1}) + \sin(\varphi_{2}) \right\rangle
\end{equation}
where $I_{c}$ is the critical current of each JJ, $\varphi_{1,2}$ are the gauge-invariant phases across the JJs, and the expectation value is taken with respect to the ground state of the CPT Hamiltonian (Eq. \ref{eq:CPT_hamiltonian}).  To evaluate this expectation value, we first rewrite the above expression in terms of $\bar{\varphi} = (\varphi_{1}+\varphi_{2})/2$ and $\delta\varphi = (\varphi_{1}-\varphi_{2})/2$ according to
\begin{equation}
I_{\mrm{circ}} = I_{c}\sin(\bar{\varphi})\left\langle \cos(\delta\varphi) \right\rangle
\end{equation}
where the average phase $\bar{\varphi}$ is independent of the state of the CPT.  The phase difference $\delta\varphi$, on the other hand, is an operator that is conjugate to the number of excess Cooper pairs $N$ on the island \cite{thesis_cottet}.  Using the canonical commutation relation between these operators, the circulating current can be written
\begin{equation}
I_{\mrm{circ}} = \frac{1}{2}I_{c}\sin(\bar{\varphi})\left\langle \sum\limits_{N\in\mathbb{Z}}\Bigl(\ketbra{N+1}{N}+\ketbra{N}{N+1}\Bigr) \right\rangle.
\end{equation}

This expectation value can now be evaluated using the charge-state truncation method discussed in Section \ref{sec:cCPT_Hamiltonian}.  We provide an upper-bound on this circulating current by setting $\sin(\bar{\varphi})=1$ and finding the maximum expectation value over the accessible range of gates $-0.65\leq n_{g} \leq 0.65$, which yields $|I_{\mrm{circ}}| \lesssim I_{c}/5$.  Since the critical current of each JJ is given by $I_{c} = 2\pi E_{J}/\Phi_{0} \approx 30$ nA, the magnitude of the circulating current is bounded by $|I_{\mrm{circ}}| \lesssim 6$ nA.  Using our above estimate for the self-inductance of the loop, we find $L_{\mrm{loop}}|I_{\mrm{circ}}|/\Phi_{0} \lesssim 0.03 \ll 1$, such that we are justified in neglecting the effect of the SQUID loop's self-inductance.

\begin{figure}[!t]
\includegraphics{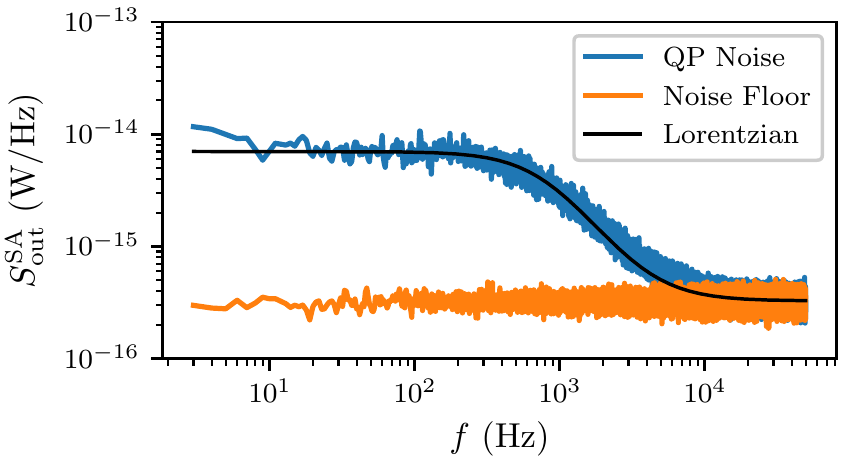}
  \caption{Output power spectral density at $f_{0}+f$ due to frequency fluctuations induced by quasiparticle poisoning.}
  \label{fig:quasiparticle_noise_spectrum}
\end{figure}

\section{Power Spectrum of Quasiparticle Poisoning}
\label{sec:power_spectrum_of_quasiparticle_poisoning}

As discussed in Section \ref{sec:tunable_resonant_frequency}, quasiparticle poisoning (QP) occurs when the cCPT is biased near $n_{g}=0.7$.  This manifests itself as random switching between two different resonant frequencies as quasiparticles tunnel onto and off of the CPT island.  When measured with a VNA, this gives rise to two visible resonances as in Fig. \ref{fig:multiperiod_f0}(b).  Using the same techniques as in Section \ref{sec:power_spectra_of_frequency_fluctuations}, however, we can also measure the power spectrum of these switching events.

To do so, we bias the cCPT to $(n_{g}, \Phi_{\mrm{ext}})=(0.7, 0.0)$.  The only sources of frequency fluctuations at this point are charge noise and the random switching due to QP (ignoring the effect of quantum fluctuations as in Section \ref{sec:power_spectra_of_frequency_fluctuations} since they cannot be resolved with this measurement).  Since the odd-parity resonance is effectively at $n_{g}=-0.3$ its resonant frequency is not very sensitive to gate noise, so the frequency fluctuations are dominated by the switching due to QP.  We therefore put a carrier in at the odd-parity resonant frequency ($f_{0} = 5.789$ GHz) and measure the output power spectral density near the reflected carrier, which encodes the power spectral density of frequency fluctuations.  This measurement is performed with all the same parameters as those discussed in Section \ref{sec:power_spectra_of_frequency_fluctuations}.  

The results of this measurement are shown in Fig. \ref{fig:quasiparticle_noise_spectrum}.  We find that the output power spectral density due to QP is well-modeled by a Lorentzian, as we expect for the random telegraph signal associated with QP \cite{Sun2012,Riste2013}.  The best fit of the excess noise to a Lorentzian is 
\begin{equation}
S_{\mrm{out}}^{\mrm{SA}}(f_{0}\pm f) = \frac{6.7\times 10^{-15}}{1+(f/830~\mrm{Hz})^{2}}~\mathrm{W/Hz}.
\end{equation}
This corner frequency of $830$ Hz is consistent with other measurements of QP reported in the literature \cite{Sun2012,Riste2013}.


%

\end{document}